\documentclass[useAMS,usenatbib,epsfig, a4paper]{mn2e}
\usepackage{epsfig}
\usepackage{times}
\usepackage{amssymb}
\usepackage{amsmath}

\newcommand{\ba}{\begin{eqnarray}}
\newcommand{\ea}{\end{eqnarray}}

\def\gtorder{\mathrel{\raise.3ex\hbox{$>$}\mkern-14mu
             \lower0.6ex\hbox{$\sim$}}}
\def\ltorder{\mathrel{\raise.3ex\hbox{$<$}\mkern-14mu
	                  \lower0.6ex\hbox{$\sim$}}}

\title[Fast and precise map-making for massively multi-detector CMB experiments]{Fast and precise map-making for massively multi-detector CMB experiments}
\author[D.~Sutton, J.A.~Zuntz and P.G.~Ferreira et al.]{D.~Sutton $^1$, J.A.~Zuntz $^1$, P.G.~Ferreira $^1$, M.L.~Brown $^2$, H.K.~Eriksen $^{3,4}$,~B.R. Johnson $^5$, \newauthor A.~Kusaka $^6$, S.K. N{\ae}ss $^3$ and I.K.~Wehus $^7$\\ 
$^1$ Oxford Astrophysics, University of Oxford, Denys Wilkinson Building, 1 Keble 
Road, Oxford OX1 3RH, United Kingdom\\
$^2$ Kavli Institute for Cosmology Cambridge, Madingley Road, Cambridge CB3 OHA, UK\\ 
$^3$ Institute of Theoretical Astrophysics, University of Oslo, P.O. Box 1029 Blindern, N-0315 Oslo, Norway\\ 
$^4$ Center of Mathematics for Applications, University of Oslo, P.O. Box 1053 Blindern, N-0316 Oslo, Norway\\
$^5$ University of California, Department of Physics, 279 LeConte Hall, Berkeley, CA 94720\\
$^6$ The University of Chicago, The Kavli Institute for Cosmological Physics, 933 East 56th Street, Chicago, IL 60637 \\
$^{7}$ Department of Physics, University of Oslo, P.O. Box 1048 Blindern, N-0316 Oslo, Norway}
\date{\today}
\pubyear{2009}

\begin{document}

\maketitle

\begin{abstract}
Future cosmic microwave background (CMB) polarisation experiments aim to measure an unprecedentedly small signal - the primordial gravity wave component of the polarisation field B-mode. To achieve this, they will analyse huge datasets, involving years worth of time-ordered data (TOD) from massively multi-detector focal planes.  This creates the need for fast and precise methods to complement the M-L approach in analysis pipelines.
In this paper, we investigate fast map-making methods as applied to long duration, massively multi-detector, ground-based experiments, in the context of the search for B-modes.  We focus on two alternative map-making approaches: destriping and TOD filtering, comparing their performance on simulated multi-detector polarisation data.
We have written an optimised, parallel destriping code, the DEStriping CARTographer (\textsc{descart}), that is generalised for massive focal planes, including the potential effect of cross-correlated TOD $1/f$ noise.  We also determine the scaling of computing time for destriping as applied to a simulated full-season data-set for a realistic experiment.
We find that destriping can out-perform filtering in estimating both the large-scale E and B-mode angular power spectra.  In particular, filtering can produce significant spurious B-mode power via EB mixing.  Whilst this can be removed, it contributes to the variance of B-mode bandpower estimates at scales near the primordial B-mode peak. For the experimental configuration we simulate, this has an effect on the possible detection significance for primordial B-modes.  
Destriping is a viable alternative fast method to the full M-L approach that does not cause the problems associated with filtering, and is flexible enough to fit into both M-L and Monte-Carlo pseudo-$C_{\ell}$ pipelines.

\end{abstract}

\begin{keywords}
cosmic microwave background -- methods: data analysis -- methods: statistical
\end{keywords}

\section{Introduction}

The temperature anisotropies of the cosmic microwave background radiation (CMB), and recently the polarisation anisotropies, have been used to constrain a set of cosmological parameters to establish a standard model of the universe (\citealt{smoot:1992}, \citealt{hanany:2000}, \citealt{melchiorri:2000}, \citealt{kovac:2002}, \citealt{spergel:2003}, \citealt{hinshaw:2009}, \citealt{brown:2009b}, \citealt{chiang:2009}).  The emphasis has now moved from revealing the contents and structure of the universe to explaining the unknown physics that generated them.  Underpinning the standard model is the paradigm of inflation, which hypothesises a period of super-luminal expansion at very early times and identifies Gaussian, scale-free quantum fluctuations as the seeds of large-scale structure.  Inflation models are predicted to uniquely generate tensor perturbations to the metric in the form of a stochastic gravitational-wave background that generates a divergence-free ``B-mode" polarisation pattern in the CMB that is not generated by scalar perturbations (\citealt{seljak:zaldarriaga:1997}; \citealt{kamionkowski:1997}).  The amplitude of the B-mode signal is inflation model dependent but is generally predicted to be orders of magnitude smaller than the temperature anisotropy.

In order to detect such a subtle signal, we must build experiments of much greater sensitivity than has been possible so far.  Due to the physical limits on the sensitivity of individual detectors, the next generation of CMB polarisation experiments aim to achieve this by using very large arrays of detectors.  For example, the Q/U Imaging ExperimenT (\textsc{quiet}) is operating  in its first phase with 91 detector horns, each of which produces 4 independent measurements (2 of Q and 2 of U), and is planned to move to a 1500 horn arrangement in its second phase. Other examples include the balloon-born \textsc{spider}, which will operate with 1024 detectors in its largest band \citep{mactavish:2008}, \textsc{ebex} \citep{oxley:2005} which will use 1440 detectors over three bands and the ground based \textsc{polarbear} \citep{lee:2008}, which will use a total of 1274 detectors. The trend is the same for high-resolution temperature anisotropy experiments, such as APEX-SZ, which is operating with 330 detectors \citep{halverson:2009}, the Atacama Cosmology Telescope (\textsc{act}), which has begun operating with $1024$ detectors per band \citep{kosowsky:2003, fowler:2007}, and the South Pole Telescope, which has 465 in its largest band \citep{plagge:2009}.

To detect B-modes using the new generation of experiments, tight control of experimental systematics will be required.  One of these systematics is correlated noise.  Detectors display long term noise drifts that are characterised by a $1/f$ power spectrum, resulting in correlated noise in the time-streams.  In addition to this, ground-based experiments see an even larger $1/f$ noise sourced from atmospheric fluctuations, which may have a polarised element \citep{hanany:2003}.  These noise sources can also be correlated between detector time-streams, for example through detector read-out electronics and due to the spatial correlation of atmospheric fluctuations projecting onto the focal plane.

It is requisite of data analysis methods to characterise and reduce the effects of noise and systematic errors.  The pipeline involves a number of steps of radical data compression.  The raw output from the telescope detectors is time-ordered data (TOD), a massive dataset that contains signal varying with the telecope pointing information.  TOD is compressed into a dataset of manageable size and in the natural form for a sky varying signal - a sky map.  From the sky map, we estimate the set of angular power spectra, including the decomposition of the polarisation field into E and B-mode amplitudes (accounting for spatial systematics such as foregrounds), from which we estimate cosmological parameters. The best place to remove correlated noise and time-variant systematic errors is in the map-making step.

The two principle approaches to map-making for CMB experiments are the maximum-likelihood  (M-L) methods (\citealt{tegmark:1997a}, \citealt{dore:2001}, \citealt{stompor:2002}, \citealt{de-gasperis:2005}), which produce optimal maps slowly, and TOD filtering, which is used with Monte-Carlo pseudo-C$_{\ell}$ methods (\citealt{hivon:2002}, \citealt{szapudi:2001}) and is very fast but necessarily sub-optimal.

The M-L algorithms produce optimal maps in that they minimise the noise whilst completely preserving the signal.  The whole pipeline from time-ordered data to power-spectrum can be optimal and maximum-likelihood, including TOD noise estimation \citep{ferreira:jaffe:2000} and power-spectrum estimation (\citealt{tegmark:1997b}, \citealt{borrill:1999}), the latter of which uses the analytic pixel-noise covariance matrix.  This approach has been used very successfully and can be extended to include information on systematics \citep{stompor:2002}.  
Contemporary M-L codes, such as \textsc{roma} \citep{de-gasperis:2005} and \textsc{mapcumba} \citep{dore:2001}, do not require explicit evaluation of the pixel-noise covariance matrix or its inverse (known as the ``weight matrix"), and have been convincingly demonstrated as viable options for single-detector Planck data analysis including polarisation (\citealt{ashdown:2009}, and references therein).  A M-L code that includes detector correlations, called \textsc{sanepic}, has recently been developed and successfully applied to the BLAST dataset \citep{patanchon:2008}, a short duration balloon experiment involving hundreds of detectors.

As future data-sets become very large, the M-L algorithm becomes increasingly difficult to implement.  The method may be tenable for some massively multi-detector experiments, but will require massive and expensive computing platforms and very long computation times.  The later stages of CMB data analysis require information on noise power and correlation and the propagation of systematics in the maps that is effectively supplied by passing hundreds of Monte-Carlo simulations through the map-making pipeline (eg: \citealt{hivon:2002, mactavish:2008, brown:2009, takahashi:2009}).  The M-L algorithm is not suited for such tasks, creating the need for fast methods that can be used on medium-sized computing platforms and used over and over again.  Such methods could be used to complement the M-L algorithm in analysis pipelines, or even to replace it if their performances become close to optimal.

The common alternative to M-L map-making is to apply a high-pass pre-whitening filter to the TOD.  The noise part of filtered TOD is uncorrelated, so a map with uncorrelated noise can be returned by averaging the filtered TOD corresponding to each pixel (a naive map).  This fast method is very well suited to Monte-Carlo simulations, but critically distorts the signal part of the TOD.  The \textsc{master} method \citep{hivon:2002} mitigates this effect by evaluating a filter transfer function from Monte-Carlo signal-only realisations that can be deconvolved in multipole space.  This approach is non-optimal and introduces extra variance into the power spectrum estimates, especially at low-$\ell$ where the primordial B-mode signal is expected. 

The destriping method is being developed as a ``third way" to CMB map-making (\citealt{burigana:1997}, \citealt{delabrouille:1998}, \citealt{maino:1999}, \citealt{keihanen:2004}, \citealt{keihanen:2005}, \citealt{kurki-suonio:2009}, \citealt{keihanen:2009}).  Destriping pre-whitens the TOD by approximating the low frequency noise part as a series of offset functions, which are  then subtracted from the TOD.  Long baseline noise drifts from $1/f$ noise are subtracted without filtering the signal part of the TOD.  The method is faster than M-L map-making and is tuneable in that the offset function length can be varied, resulting in fast and dirty maps or slow and near optimal maps.  Most destriping codes exploit the great-circle scanning strategy of satellite experiments to subtract long offset functions that are averages of the TOD in one re-pointing (see eg: \citealt{ashdown:2007a} for a discussion).  The \textsc{madam} code \citep{keihanen:2005}, was the first to incorporate noise information, through a prior on the offset function amplitude distribution that permits the use of short offset amplitudes.  A variant of destriping that does not use noise information has recently been applied to total-intensity data from ACT to produce Sunyaev-Zeldovich (SZ) effect maps of galaxy clusters, using destriping baselines to model the inter-detector common-mode atmospheric noise \citep{hincks:2009}.

In a previous paper \citep{sutton:2009}, we showed that destriping with short baselines using noise information is near-optimal when applied to non-circular scanning strategies, such as the cross-linked, azimuth only scans of ground-based experiments.  In this paper, we extend the destriping algorithm to include information on noise correlations between detectors and investigate the performance of our new code, the DEStriping CARTographer (\textsc{descart}), as applied to multi-detector TOD simulations.  For these simulations, the Q and U signals are not modulated to the high-frequency white noise part of the spectrum, as is the case for half-wave plate modulation experiments, and are contaminated with $1/f$ noise.  We compare it to the \textsc{master} method with filtering, the established fast map-making method, to determine the strengths and limitations of the filtering approach for realistic future data-sets.

Given the importance of computation time for large data-sets, we also investigate the algorithmic scaling of \textsc{descart} as applied to a large full-season multidetector data-set.  We concentrate on polarisation measurements for B-mode experiments.  The \textsc{descart} algorithm is also applicable to multi-detector SZ experiments, which we plan to investigate in a future paper.

This paper is organised as follows: in \S\ref{destriping and filtering section}, we review map-making formalism of filtering and destriping in the context of mult-detector polarisation data; in \S\ref{comparison of the pipelines section}, we directly compare the filtering and destriping pipelines through simulations for a range of noise scenarios, comparing errors in timestreams, maps and E and B-mode angular power spectra; and in \S\ref{quiet simulations section}, we apply \textsc{descart} to a much larger simulation of a massively multi-detector experiment, focussing on algorithmic scalings and the partitioning of the analysis.

\section{Destriping and filtering} \label{destriping and filtering section}

\subsection{The map-making problem}

The map-making problem requires a solution for a sky map, $x_{p}$, given detector time-streams $d_{t}$, pointing information and possibly noise information.  The time-ordered data (TOD) is modelled as
\begin{equation}
d_{t}= A_{tp} x_{p} + n_{t}
\end{equation}
where $A_{tp}$ is a pointing matrix that contains all of the relevant pointing information, $t$ indexes time and $p$ indexes sky pixel.  The noise time-stream $n_{t}$ is commonly assumed to be Gaussian and piece-wise stationary, satisfying
\begin{eqnarray}
\langle n_{t} \rangle = 0 \\
\langle n_{t} n_{t'} \rangle = C_{N}
\label{tod covariance}
\end{eqnarray}
In general, $n_{t}$ is correlated noise, such that (\ref{tod covariance}) is not diagonal.  If the noise is stationary, its covariance (\ref{tod covariance}) is well described as symmetric Toeplitz, or as a function of separation $C_{Ntt'}= f(t-t')$. If the instrumental beams are symmetric, the pointing matrix for a temperature measurement $A_{tp}$ is composed of 1s and 0s and contains only one non-zero element per time row indicating which sky pixel the beam centre falls upon.

This basic framework can be extended to multiple detectors and to measurements of polarisation.  In the experimental configuration we simulate in this paper, each detector ``pixel" makes direct measurements of Q and U Stokes parameters in its own frame of reference.  The rotation from sky to detector frame of reference is subsumed into the pointing matrix
\begin{eqnarray}
P_{tp}= R(\theta_{t}) A_{tp} ,\\
R(\theta_{t}) = \left( \begin{array}{cc}
\cos{2\theta_{t}} & \sin{2\theta_{t}} \\
-\sin{2\theta_{t}} & \cos{2\theta_{t}}
\end{array} \right).
\label{pol pointing matrix}
\end{eqnarray}

For simplicity in notation, we define the pointing matrix for detector-Q time-streams as $P_{tp}= (\cos{2\theta_{t}}, \sin{2\theta_{t}}) A_{tp}$ and for detector-U time-streams as $P_{tp}= (-\sin{2\theta_{t}}, \cos{2\theta_{t}}) A_{tp}$.  From this point we drop the distinction between them and consider $P^{k}_{tp}$ as the appropriate pointing matrix for the time-stream indexed by $k$.  The map on which the polarisation pointing matrices operate contains maps of the Q and U Stokes parameters, concatenated end-to-end $\vec{x}= (x^{Q}_{p} , x^{U}_{p})$.

For multiple time-streams, we stack the polarisation pointing matrices, data and noise time-streams end-to-end to produce the multi-detector analogue to the data model,
\begin{eqnarray}
\vec{d} = \mathbf{P} \vec{x} + \vec{n} \\
\vec{d}= \left( \begin{array}{c} d^{1}_{t} \\ d^{2}_{t} \\ \vdots \\ d^{n}_{t} \end{array} \right),
\vec{n}= \left( \begin{array}{c} n^{1}_{t} \\ n^{2}_{t} \\ \vdots \\ n^{n}_{t} \end{array} \right),
\mathbf{P}= \left( \begin{array}{c} P^{1}_{tp} \\ P^{2}_{tp} \\ \vdots \\ P^{n}_{tp} \end{array} \right).
\end{eqnarray}

Given the data model, we can construct estimators for the sky map $\vec{x}$.  The M-L solution to the map-making problem is 
\begin{equation}
\vec{x}= (\mathbf{P^{T} C_{N}^{-1} P})^{-1} \mathbf{P^{T} C_{N}^{-1}} \vec{d},
\end{equation}
where $\mathbf{P^{T}}$ is the transpose of $\mathbf{P}$ and has the effect of summing TOD into a map, rather than projecting a map onto TOD.  In the limit that we consider the TOD noise to be white (uncorrelated), the solution becomes very simple, corresponding to naive binning and averaging,
\begin{equation}
\vec{x}= (\mathbf{P^{T} P})^{-1} \mathbf{P^{T}} \vec{d}.
\end{equation}

\subsection{The destriping solution}

The maximum-likelihood method can be approximated by the destriping algorithm (\citealt{sutton:2009}, \citealt{ashdown:2009} and references therein).  The noise vector $\vec{n}$ is modelled as being composed of two components: uncorrelated (or \emph{white}) noise;  and a series of discrete offset functions that represent all of the correlated noise.

With $n^{W}_{t}$ representing the random Gaussian realisation of white noise, the new data model for a single detector is
\begin{equation}
d_{t} = P_{tp} x_{p} + n^{W}_{t} + F_{t\alpha} a_{\alpha}
\end{equation}
where the matrix $F_{t\alpha}$ maps a set of basis functions, with amplitudes $a_{\alpha}$, onto the time-stream.  

For multi-detector systems, we generalise the basis function amplitude vectors by concatenating them into a vector $\vec{a}= (a^{1}_{\alpha}, a^{2}_{\alpha}, \dots, a^{n}_{\alpha})$, and defining a block-diagonal multi-detector offset pointing matrix
\begin{equation}
\mathbf{F} = \left(
\begin{array}{ccc}
F^{1}_{t\alpha}  & 0 & \ldots  \\
0 & F^{2}_{t\alpha}  & \ldots \\
\vdots & \vdots & \ddots  
\end{array} \right)
\end{equation}
The simplest choice of basis function is a constant offset
\begin{equation}
\mathbf{F_{t\alpha}} = \left\{ \begin{array}{cc}
1 & \textrm{$t \in \Delta_{\alpha}$} \\
0 & \textrm{otherwise,}
\end{array} \right.
\label{const_chunk}
\end{equation}
where $\Delta_{\alpha}$ is a chunk of the time-stream with length $\lambda_{\textrm{d}}$.  With this definition, $F_{t\alpha}a_{\alpha}$ becomes a vector of constant offsets, with amplitudes $\vec{a}$, approximating the correlated noise. More complicated basis functions can be chosen, for example Fourier series or Legendre polynomials \citep{keihanen:2004}, though we focus on using short uniform baselines.

The amplitudes $\vec{a}$ will be Gaussian random numbers satisfying
\begin{eqnarray}
\langle a_{\alpha} \rangle &=& 0 \\
\langle a_{\alpha} a^{T}_{\alpha'} \rangle &=& \mathbf{C_{a}}_{\alpha \alpha '}
\label{a Ca constraints}.
\end{eqnarray}
With $\vec{a}$ as a second set of parameters, the posterior for both the map and the offset amplitudes is 
\begin{equation}
P(\vec{x},\vec{a}|\vec{d}) \propto P(\vec{d}|\vec{x},\vec{a}) P(\vec{x}) P(\vec{a})
\label{destr_bayes},
\end{equation}
where we have used the fact that the CMB map and the offset amplitudes are completely independent.  We do not wish to assume any prior for the CMB map \footnote{It is standard practice to assume priors on cosmological parameters, but the sky map is our first estimator of signal and should be free of map priors, which themselves have a very non-linear dependence on the cosmological parameter set.}, so we consider $P(\vec{x})$ to be constant.

The next step is to decide what prior information on the amplitudes to include through $P(\vec{a})$.  If we have an estimate of the power spectrum of the correlated noise, which the offset functions are designed to approximate, we can include prior information through the offset covariance matrix $\mathbf{C_{a}}$.  The probability distribution for the offsets is Gaussian
\begin{equation}
P(\vec{a}) = 
|\mathbf{C_{a}}|^{-1/2} \exp{\bigg(-\frac{1}{2} \vec{a}^{T} \mathbf{C_{a}^{-1}} \vec{a}\bigg)}
\label{offsets prior},
\end{equation}
where $\mathbf{C_{a}}$ is the covariance of the offset amplitude estimates (ignoring factors of $2\pi$).  If we have prior knowledge of $\mathbf{C_{N}}$, we can build a prior $\mathbf{C_{a}}$ through 
\begin{equation}
\mathbf{C_{a}}= (\mathbf{F^{T}F})^{-1} \mathbf{F^{T} C_{N} F} (\mathbf{F^{T}F})^{-1} .  
\label{definition of prior}
\end{equation}

The joint posterior is a product of two Gaussians: the likelihood of the data and the prior for the amplitude estimates.  We can maximise the posterior by minimising the function $f$, where
\begin{eqnarray}
f &=& -2 \ln| P(\vec{x},\vec{a}|\vec{d})| \nonumber \\
 & = &(\vec{d}- \mathbf{F}\vec{a}-\mathbf{P}\vec{x})^{T} \mathbf{C_{W}}^{-1} (\vec{d}- \mathbf{F}\vec{a}-\mathbf{P}\vec{x}) \nonumber \\
 &&+ \vec{a}^{T} \mathbf{C_{a}^{-1}} \vec{a}
 \label{simple destriping chisq}.
 \end{eqnarray}
Solving $\partial f / \partial \vec{x}=0$, we obtain an expression for the map
\begin{equation}
\vec{x}= (\mathbf{P^{T} C_{W}^{-1} P})^{-1} \mathbf{P^{T} C_{W}^{-1}} (\vec{d}-\mathbf{F}\vec{a})
\label{destriped_map}.
\end{equation}
This result can be understood as follows: The correlated noise $\mathbf{F}\vec{a}$ is subtracted from time-stream $\vec{d}$, leaving a time-stream composed of only signal and white noise, which is naively averaged into a map (the maximum-likelihood map for a time-stream with uncorrelated noise is the pixel average).

To determine the offsets, we begin by substituting (\ref{destriped_map}) into (\ref{simple destriping chisq}) and gathering all the terms involving the pointing matrix $\mathbf{P}$ into a new operator, $\mathbf{Z}$,
\begin{equation}
f = (\vec{d} - \mathbf{F} \vec{a})^{T} \mathbf{Z^{T} C_{W}^{-1} Z} (\vec{d} - \mathbf{F}\vec{a})+ \vec{a}^{T} \mathbf{C_{a}^{-1}} \vec{a}
\label{simple destriping chisq with Z}, 
\end{equation}
where
\begin{equation}
\mathbf{Z} = \mathbf{I} - \mathbf{P(P^{T} C_{W}^{-1} P)^{-1} P^{T} C_{W}^{-1}}
\label{definition of Z}
\end{equation}
and $\mathbf{I}$ is the identity matrix.  $\mathbf{Z}$ is a signal cleaning operator that estimates the naive map from the timestream (including both signal and any noise that is indistinguishable from signal) and removes it.

Solving $\partial f / \partial \vec{a} = 0$, we obtain an estimator for the offset amplitudes
\begin{equation}
(\mathbf{F^{T}C_{W}^{-1}ZF + C_{a}^{-1}})\vec{a} = \mathbf{F^{T} C_{W}^{-1}Z}\vec{d} 
\label{amp_destr}.
\end{equation}

This is an inverse problem like that of the maximum likelihood algorithm.  The system can be solved quickly, using the preconditioned conjugate gradients method, at a fraction of the processing time required to solve for the full maximum-likelihood map.  An effective preconditioner $\mathbf{K}$, can be constructed and applied easily from the offset prior covariance matrix
\begin{equation}
\mathbf{K} = (\mathbf{F^{T} C_{W} F} + \mathbf{C^{-1}_{a}})
\end{equation}
Destriping solves the map-making problem by making an approximate model for the noise.  In the limit that the offset functions perfectly model the correlated noise, the map solution \emph{is} the maximum-likelihood map.

\subsubsection{Destriping correlated time-streams} \label{destriping correlated streams section}

For multi-detector systems with noise uncorrelated between time-streams, the noise covariance matrix, $\mathbf{C_{N}}$, will be non-zero only in diagonal detector-detector blocks (which themselves will be symmetric Toeplitz).  Cross-correlated noise will have non-zero off-diagonal detector-detector blocks in its $\mathbf{C_{N}}$ which will lead to non-zero diagonal block in $\mathbf{C_{a}}$ through equation (\ref{definition of prior}).

The inversion of the $\mathbf{C_{a}}$ is achieved by using some simplifying assumptions.  To a good approximation, the individual detector blocks can be considered to be circulant and can be inverted individually very easily.  A circulant matrix becomes diagonal in Fourier space and each Fourier mode $\omega$, which is independent, can be inverted as a scalar.  In the presence of non-zero off-diagonal detector blocks, each mode becomes an independent $N_{detector}^{2}$ matrix, which we invert using Cholesky decomposition. This is the same approach as is used in inverting the multi-detector noise matrix in \textsc{sanepic}: \cite{patanchon:2008}.   This entails the inversion of $n_{\mathrm{offsets}}$ matrices of dimension $n_{\mathrm{det}}$ each, once, before the iterations begin, an $\mathcal{O}(n_{\mathrm{offsets}} n_{\mathrm{det}}^3)$ operation.

The resultant matrix $(\tilde{C}_{a}^{kk'})^{-1}(\omega)$ still has diagonal detector $kk'$ blocks in Fourier space, so the matrix multiplication  $\mathbf{C_{a}^{-1}}\vec{a}$ are conducted in Fourier space to minimise the operation count
\begin{equation}
\tilde{w}_{k}(\omega) = \sum_{k'} \tilde{M}_{kk'}(\omega) \tilde{v}_{k'}(\omega)
\label{mat mult}.
\end{equation}
Here the vectors $w$ and $v$ are multi-detector vectors of offsets, and the matrix $M$ is either $\mathbf{C_{a}^{-1}}$ for the matrix application or $\mathbf{K}$ for the preconditioning step.

\subsection{The filtering approach}

Filtering approaches to map-making apply a time-domain filter to TOD in order to suppress noise or systematics in the time-stream, before naively binning the now uncorrelated timestreams.   For a filter $h(t-t')$, the map estimate is
\begin{equation}
\vec{x}= (\mathbf{P^{T} P})^{-1} \mathbf{P^{T}} h * \vec{d}
\label{filtered map}.
\end{equation}

In this paper, we consider the effect of aggressive filters intended to suppress correlated noise rather than filters specific to experimental systematics.  One choice of filter to suppress correlated noise is the prewhitening filter $h= \sigma_{W} \mathbf{C^{-1/2}_{N}}$, where $\sigma_{W}$ is the white noise standard deviation.  The covariance of the TOD filtered in this way is diagonal ($\sigma_{w}^{2} \delta_{ij}$).  Another common choice is the \emph{overwhitening} filter, $h= \sigma_{W}^{2}\mathbf{C_{N}^{-1}}$ filter, which further suppresses low frequency noise.

The filter convolution is applied in the Fourier domain, where $\mathbf{C_{N}}$ is diagonal.  This requires a Fourier transform pair to be applied to the timestream, both of which have an operational scaling of $\mathcal{O}(N_{t} \log_{2} N_{\mathrm{segment}})$, where $N_{t}$ is the number of TOD and $N_{\mathrm{segment}}$ is the length of the segments over which the noise is stationary and correlated (for example the length of a constant elevation scan).

\subsection{Covariance of the estimated maps} \label{covariance section}

Both of the pipelines used in this paper construct the final map using a naive binning step.  The noise covariance of maps constructed using naive binning is 
\begin{equation}
\mathbf{C}= (\mathbf{P^{T} P})^{-1} \mathbf{P^{T} C^{TOD}_{N} P} (\mathbf{P^{T} P})^{-1},
\label{naive map covariance}
\end{equation}
where $\mathbf{C^{TOD}_{N}}$ is the covariance of the TOD.  For the filtering pipeline, 
$\mathbf{C^{TOD}_{N}}= \int e^{it\omega} h C_{N} h^{\dagger} d\omega$, which for the prewhitening filter is diagonal and produces a diagonal pixel covariance.

In the destriping pipeline, two sets of parameters are solved for; the map and the offset amplitudes.  The Fisher matrix for the full parameter set is
\begin{eqnarray}
\begin{array}{ccc}
\mathcal{F} &=& \left(
\begin{array}{cc}
 \partial^{2} f / \partial x^{2} &  \partial^{2} f / \partial x\partial a  \\
  \partial^{2} f / \partial a\partial x  &   \partial^{2} f / \partial a^{2}
\end{array}
\right) \\
&=& \left( \begin{array}{cc}
\mathbf{P^{T} C_{W}^{-1} P}   &  \mathbf{F^{T} C_{W}^{-1} P} \\
\mathbf{P^{T} C_{W}^{-1} F}   &  \mathbf{F^{T} C_{W}^{-1} F} + \mathbf{C_{a}^{-1}}
\end{array} \right)
\end{array}
\label{destriping Fisher},
\end{eqnarray}
where $f$ is the posterior in (\ref{simple destriping chisq}).  Inverting this matrix by partition gives the noise covariance of the destriped maps
\begin{flushleft}
\begin{eqnarray}
\mathbf{C}& = & \big( \mathbf{P^{T} C_{W}^{-1} P} - \label{destriping covariance}\\
& &(\mathbf{P^{T} C_{W}^{-1} F} )( \mathbf{F^{T} C_{W}^{-1} F} + \mathbf{C_{a}^{-1}})^{-1}(\mathbf{F^{T} C_{W}^{-1} P} )  \big)^{-1}, \nonumber
\label{covariance matrix}
\end{eqnarray}
\end{flushleft}
which can be solved for low-resolution maps of large angular scales \citep{keskitalo:2009}.

This solution assumes that the destriping data model is correct - that the correlated noise is perfectly described by the offset functions.  Whilst this is an acceptable approximation for short offset function lengths, it breaks down for long baseline functions.  In this regime, unmodelled correlated noise contributes significantly to the map covariance.

The quantity that destriping tries to measure is the average of the noise in a segment of the TOD, often referred to as the ``reference baseline", defined by $r= (F^{T} F)^{-1} F^{T} d_{n}$ where $d_{n}$ is noise-only TOD.  The covariance described by equation (\ref{destriping covariance}) is from the error between the baseline amplitudes estimated by (\ref{amp_destr}) and these reference baselines.  If this error is uncorrelated with the unmodelled correlated noise within a TOD segment, as it is for our simulations (see \S \ref{timestreams and maps section}), we can account for the unmodelled noise by adding a correction term $\mathbf{C^{corr}}$ to the covariance matrix (\ref{destriping covariance}),

\begin{eqnarray}
\mathbf{C^{corr}} &=&   (\mathbf{P^{T} P})^{-1} \mathbf{P^{T} C^{des}_{N} P} (\mathbf{P^{T} P})^{-1} 
\label{matrix simplification 1}\\
\mathbf{C_{N}^{des}} &=& \mathbf{C_{N}} + \mathbf{F C_{a} F^{T}} \nonumber \\
&&- \mathbf{C_{N} F (F^{T}F)^{-1} F^{T}  + F(F^{T} F)^{-1} F^{T} C_{N}},
\label{matrix simplifications}
\end{eqnarray}
where $\mathbf{C_{a}}$ is the offset prior and $\mathbf{C_{N}}$ is the TOD noise covariance. For the symmetric Toeplitz $\mathbf{C_{N}}$ and $\mathbf{C_{a}}$ we consider, $\mathbf{C_{N}^{des}}$ is also symmetric Toeplitz.  The implementation of (\ref{matrix simplification 1}) requires an $\mathcal{O}(N_{t} N_{corr})$ summation, where $N_{corr}$ is the correlation length of the reference-offset subtracted timestream.  Any correlation in this timestream is due to unmodelled $1/f$ and has a short correlation length.

\subsection{The code} \label{code section}

The \textsc{descart} code in which we implemented the above algorithm is 
written in Fortran and uses MPI for communication; this ensures 
scalability to the largest available systems.  The communication 
operations occur when generating and distributing the collective naive 
map for the current iteration, and when correlating horns in a scan.

We minimize the memory usage of the code with a juggling of the 
data.  When all the timestreams that might be correlated (those within 
a single scan) are loaded they are immediately used in two ways: added 
to an accumulated naive map, and reduced into baseline offsets.  They 
are then discarded.  The application of the prior and likelihood 
matrix operations during the PCG requires only a single allocated 
one-detector timestream at a time (per process).  Finally, when the best-fit 
offsets have been computed, they are converted directly into a map 
(which represents the correlated noise in the original naive map).  
Deducting this from the first map yields the output map.  At no point 
do we allocate full timestreams.  With this modification \textsc{descart} can 
process $900$ one-hour timestreams per 2-GB RAM process for typical baseline lengths.  This increases markedly if multiple time-streams are associated with the same horn.

As currently constituted the code is not much slower than generating 
naive maps for larger runs, since simply loading the data takes a 
significant proportion of the run time.

\section{Comparison of the pipelines} \label{comparison of the pipelines section}

In this section, we analyse the comparative performance of the filtering and destriping pipelines as applied to simulations of a constant elevation scanning multi-detector telescope.  We begin by analysing time-stream and map domain errors, the latter for which we use the root-mean-square (rms) residual as the comparative statistic.  We proceed to analysing spherical harmonic domain systematics and errors, for which we need to estimate the E and B mode angular power spectra.  For this task we use the Monte-Carlo approach of \cite{hivon:2002}, generalized for polarization power spectra as in \cite{ksmith:2006} such that estimators do not mix E and B modes.   E to B mixing is mitigated by applying an apodizing window function to the estimated maps.  The window function we use is a symmetric circular cosine apodisation around the map centre.  The apodization is the same as that used in our previous paper \citep{sutton:2009}, to which we refer the reader for details.  The method requires estimation of the power spectral noise bias $\langle N_{\ell}^{MC} \rangle$ through Monte-Carlo (MC) noise only simulations and estimation of the filter transfer function $F_{\ell}$ through signal only simulations.  Finally, we analyse the statistical significance of the B-mode detection provided by the two pipelines.

\begin{figure}
\begin{center}
\includegraphics[angle=0,width=0.48 \textwidth]{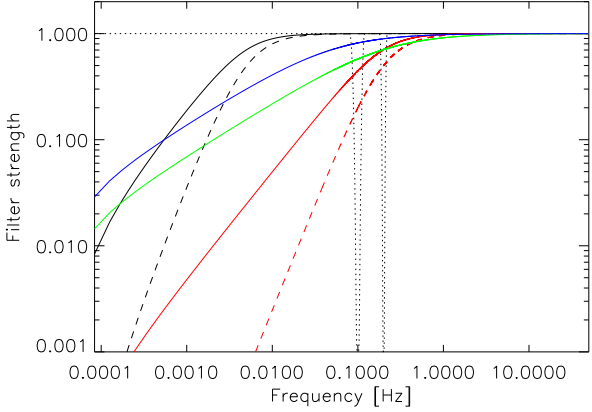}
\caption{Filters used in the filtering pipeline. The solid curves are the prewhitening filters $\mathbf{C_{N}^{-1/2}}$ and the dashed curves are overwhitening filters $\mathbf{C_{N}^{-1}}$.  Colour denotes the noise case, with black denoting $f_{k}=5$-mHz $\alpha=2$, blue $f_{k}=50$-mHz $\alpha=1$, green $f_{k}=200$-mHz $\alpha=1$, and red $f_{k}=200$-mHz $\alpha=2$. The dotted curve is the scan frequency filter, defined for the scanning frequency and its first harmonic with a bandwidth of $0.02$-Hz.}
\label{filters plot}
\end{center}
\end{figure}

Figure \ref{filters plot} shows the range of filters investigated in the filtering pipeline.  These comprise prewhitening ($\mathbf{C_{N}^{-1/2}}$) and overwhitening ($\mathbf{C_{N}^{-1}}$) filters, for low frequency noise suppression, and a scan frequency filter, used for the suppression of scan-synchronous systematics.

The destriping results in this section make use of inter-timestream noise correlation information.  The benefit of including these correlations is dependent on the experimental set-up and for the polarised dataset we simulated, the noise reduction when including the correlation was small.  For other systems, including this information may be important.  When applying the algorithm to simulated data in \S\ref{quiet simulations section}, correlation information was ignored as the experiment has negligible inter-detector correlated $1/f$.   The choice of destriping length is informed by the noise $f_{k}$: as we are incorporating prior noise information, a choice of destriping length $\lambda_{\textrm{d}} \lesssim 0.1/f_{k}$ is used to to approximate the optimal M-L map as closely as possible.

\subsection{Simulated Data-sets} \label{simulations section}
\begin{figure}
\begin{center}
\includegraphics[angle=0,width=0.4 \textwidth]{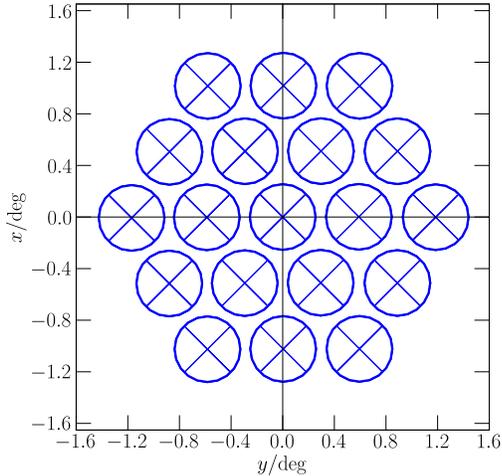}
\caption{Plan of the simulated focal plane arrangement.  Each horn produces a Q and a U time-stream in the detector's frame of reference.}
\label{hits map}
\end{center}
\end{figure}

Time-ordered data was simulated for a variety of noise cases, shown in Table \ref{parameters}, using a one-day scanning strategy comprising four constant elevation scan sets.  This scan is a typical observation of a single non-galactic field from an observatory in the Chanjnantor Scientific Reserve in Chile.  At a sampling frequency of $f_{s}=100$-Hz, the total integration time for the scans were 6 hours and 38 minutes.  We simulated an instrument with a Gaussian symmetric beam with FWHM$=12'$ and a focal plane consisting of 19 horns arranged in a hex pattern, each producing 2 time-streams.  We show this arrangement in Figure \ref{hits map}.

For each noise case, we generated 100 signal-plus-noise and 100 independent noise-only realisations.  Each realisation consisted of 38 time-streams, with Gaussian cross-correlated $1/f$ noise, Gaussian white noise, and an independent CMB signal realisation produced using \textsc{synfast} \citep{gorski:2005} from the input $C_{\ell}$s used in our previous paper (\citealt{sutton:2009}, for which the tensor to scalar ratio $r=0.1$).

Cross-correlated noise was generated by statistically correlating the $1/f$ streams.  For each time-stream, a Fourier-space Gaussian random number vector $\phi_{d}(f)$ was generated, where $d$ indexes detectors.  We built a correlation matrix $A_{dd'}= \rho_{dd'} \sigma_{d} \sigma_{d'} \sqrt{f_{k,d} f_{k,d'}}$, where $\sigma_{d}$ and $f_{k,d}$ are the noise standard deviation and knee frequency for stream $d$ respectively and $\rho_{dd'}$ is the correlation coefficient for streams $d$ and $d'$.  Correlated $1/f$ noise streams were produced by applying the 
Cholesky decomposition of this matrix to the set of $\phi_{d}$ mode-by-mode
\begin{equation}
n_{d}(f) = P^{1/2}(f) \sum_{d'} L_{dd'} \phi_{d'}(f),
\label{correlating equation}
\end{equation}
where $P(f)= (1/f)^{\alpha}$ and $A= LL^{T}$.

To get time-domain noise stream realisations with the desired auto and cross power spectra, each $n_{d}(f)$ stream was Fourier transformed and added to an independent Gaussian white noise realisation with standard deviation $\sigma_{d}$.  The white noise level itself was determined assuming a fiducial noise-equivalent Q (NEQ) of $248$-$\mu K \sqrt{s}$.

\begin{table}
\caption{Noise parameters used in the simulations.  The spectral index of unity describes the correlated noise from detectors, whilst the spectral index of two describes atmospheric noise. The ranges of knee frequency chosen are representative of frequencies experienced by low frequency ($f < 100$-GHz) experiments.  The $f_{k}$ and $\alpha$ of noise case 6 are representative of QUIET Q-band data.  The values of $\sigma_{W}$ here correspond to one year's of observing time with $1000$, $91$ and $19$ horns, where $\sigma_{W}$ has been scaled so as to produce the white noise amplitude in the final map expected for these dataset sizes.  For all cases, we simulate weakly correlated inter-detector $f^{-\alpha}$ noise, with correlation coefficient $\rho= 0.2$.}
\begin{center}
\begin{tabular}{cccc}
Parameter & White noise RMS & Knee frequency & Spectral index \\
Symbol & $\sigma_{W}$ & $f_{k}$ & $\alpha$ \\
\hline
Case 1 & 130-$\mu K$& 200-mHz& 1\\
Case 2 & 59.3-$\mu K$& 200-mHz &1 \\
Case 3 & 17.9-$\mu K$& 200-mHz& 1\\
Case 4 & 17.9-$\mu K$& 200-mHz& 2\\
Case 5 & 17.9-$\mu K$& 50-mHz& 1\\
Case 6 & 17.9-$\mu K$& 5-mHz& 2\\
\hline
\end{tabular}
\end{center}
\label{parameters}
\end{table}%

\subsection{Time-streams and maps} \label{timestreams and maps section}

\begin{figure}
\begin{center}
\includegraphics[angle=0,width=0.5 \textwidth]{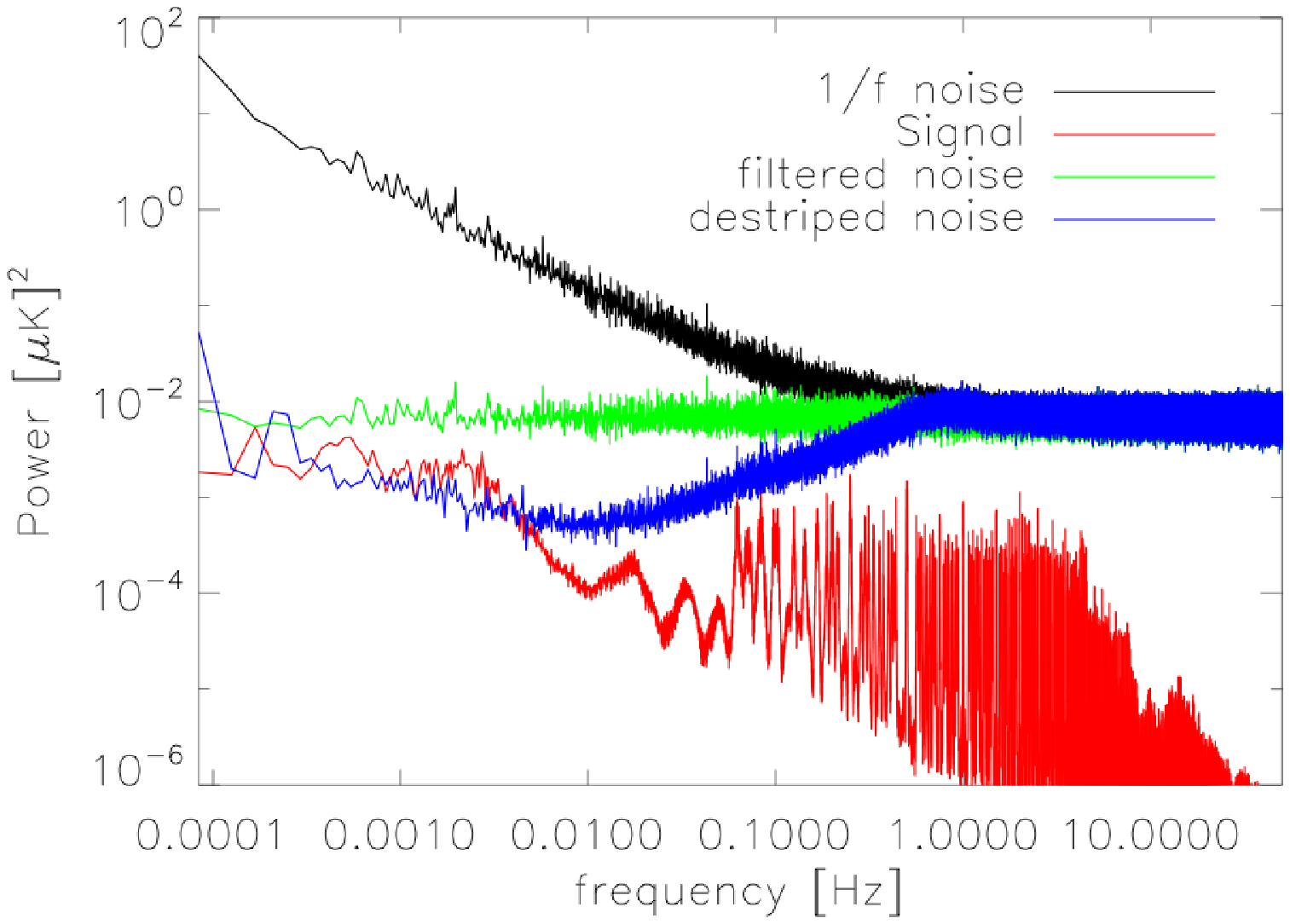} 
\includegraphics[angle=0,width=0.5 \textwidth]{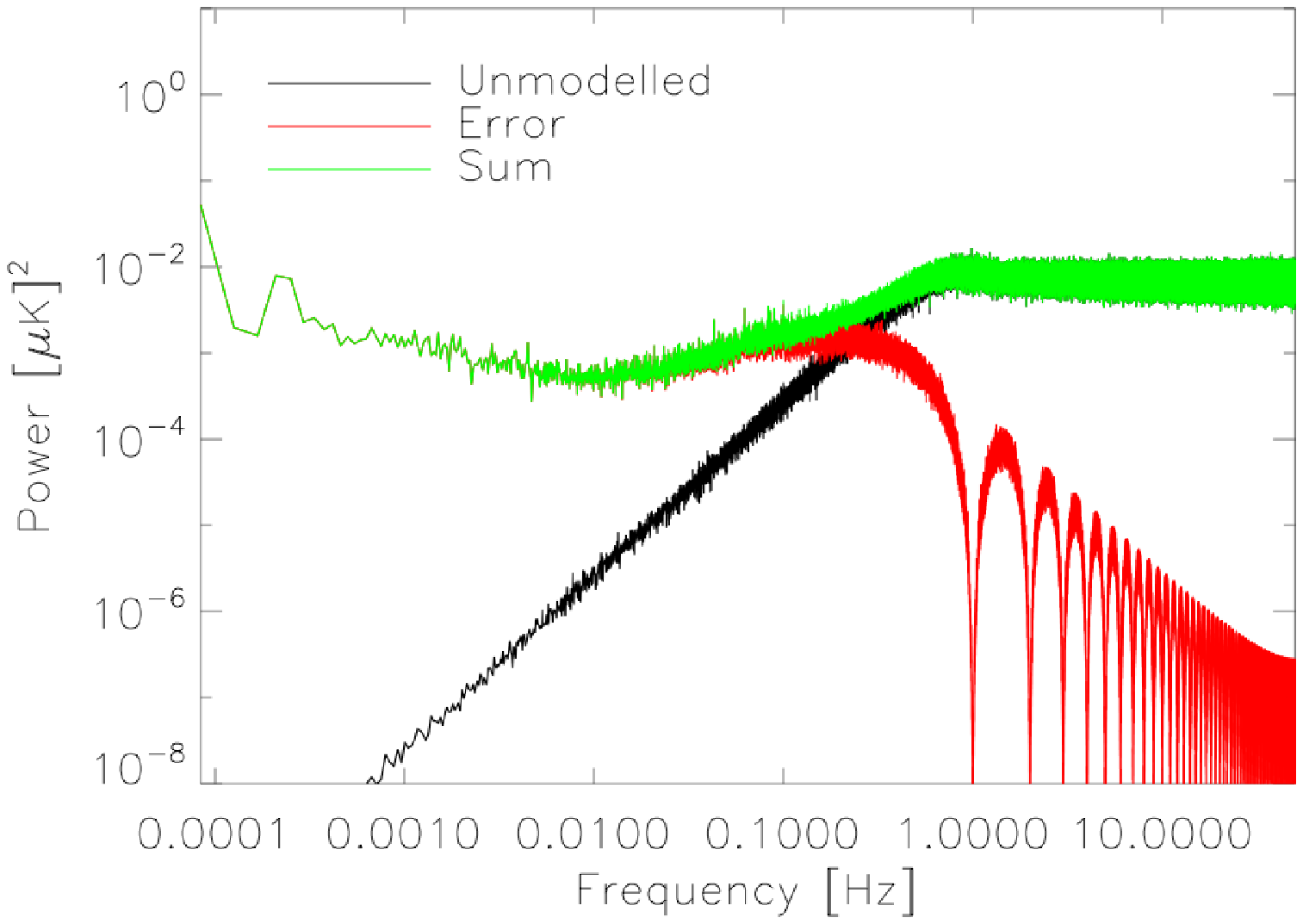}
\caption{\emph{Upper panel}: Power spectra of TOD for $1/f$ noise (black curve), filtered noise (green curve) and destriped noise (blue curve - the destriping length corresponds to $f= 1$-Hz).  Also plotted is an arbitrarily normalised power spectrum of signal only TOD (red curve). \emph{Lower panel}: Power spectrum of unmodelled noise (black curve), power spectrum of destriping error $\mathbf{F}\vec{\epsilon}$ (red curve) and the sum of these spectra.}
\label{power_plot}
\end{center}
\end{figure}

Both pipelines approach map-making by suppressing noise in the time domain before naive binning.  In the upper panel of Figure \ref{power_plot}, we show the results of using the methods on mean time-domain noise power.  The time-stream power spectrum is composed of signal (red curve) and noise (black curve).  Filtering suppresses both of these components, enforcing the flatness of the output TOD power (green curve).  Destriping only affects the noise component of the TOD, suppressing its power at frequencies smaller than the destriping offset function length (blue curve), in this case at $1$-Hz.

The noise in the destriped TOD comes from two components: unmodelled noise and destriping error.  Unmodelled noise includes all noise sources at frequencies higher than the destriping length and some power beneath it, which decreases exponentially at frequencies beneath the destriping length, behaviour well described by the application of a transfer function to the noise spectrum, as described by \citealt{kurki-suonio:2009}. Defining the reference baselines as $r= (\mathbf{F^{T}F})^{-1} \mathbf{F^{T}} \vec{d}_{n}$ (ie: the average of the noise within a baseline) the unmodelled noise is $\vec{d}_{n} - \mathbf{F} \vec{r}$.  The destriping error, defined by $\mathbf{F}(\vec{a} - \vec{r})$, describes the error in estimation of the baseline amplitude.  This error is a complicated quantity, determined by the pointing of the telescope.

The lower panel of Figure \ref{power_plot} shows the power spectrum of the unmodelled noise (black curve), the power spectrum of the destriping error (red curve) and the sum of these power spectra (green curve).  The destriping error is the dominant component of the noise in the destriped TOD at low frequencies.  At high frequencies, the unmodelled noise dominates, whilst the destriping error shows a series of troughs at multiples of the destriping length caused by the window function of the offset functions.

The sum of the power spectra is virtually identical to the power spectrum of the destriped noise TOD.  For our simulations, this validates the assumption, made in \S \ref{covariance section}, that the destriping error is uncorrelated with the unmodelled noise.  Putting the unmodelled noise $\vec{d}_{u}= \vec{d}_{n}- \mathbf{F}\vec{r}$, and the destriping error $\epsilon= \vec{a} - \vec{r}$, the power spectrum of the destriped TOD is 
\begin{eqnarray}
\int dt e^{-ift} \langle(d_{u} - F\epsilon)^{2}\rangle = \int dt e^{-ift} \Big( \langle d_{u}^{2} \rangle + \langle (F\epsilon)^{2} \rangle \nonumber \\
- \langle d_{u} \epsilon^{T} F^{T} \rangle - \langle F \epsilon d_{u}^{T} \rangle \Big)
\label{power spectra sum equation},
\end{eqnarray}
which requires the cross-correlations $\langle d_{u} \epsilon^{T} F^{T} \rangle - \langle F \epsilon d_{u}^{T} \rangle=0$.  This facilitates the use of (\ref{matrix simplifications}) to correct the destriped map covariance matrix as calculated by (\ref{covariance matrix}).

\begin{figure*}
\begin{center}
\includegraphics[angle=90,width=0.4 \textwidth]{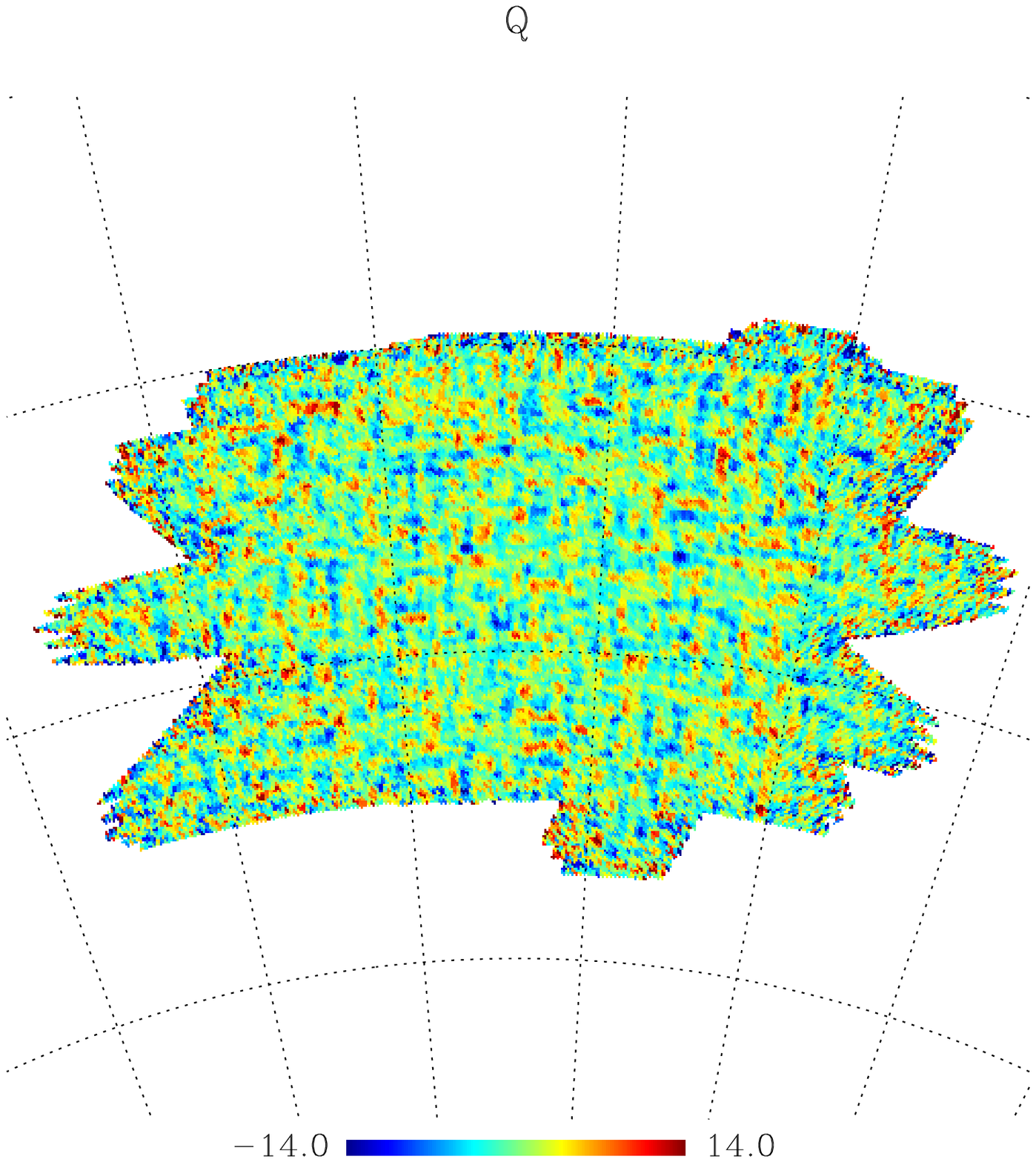}
\includegraphics[angle=90,width=0.4 \textwidth]{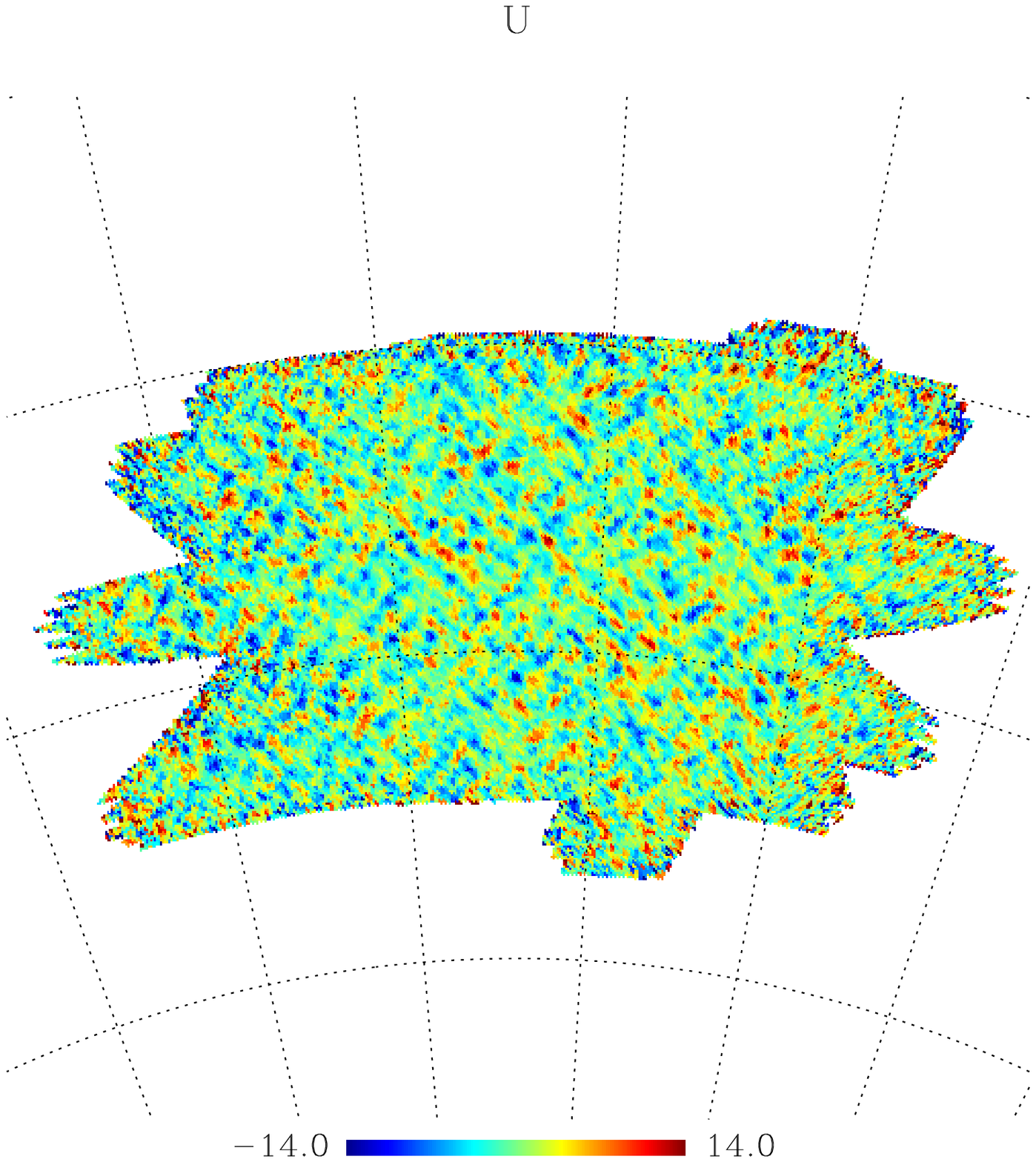} \\
\includegraphics[angle=90,width=0.4 \textwidth]{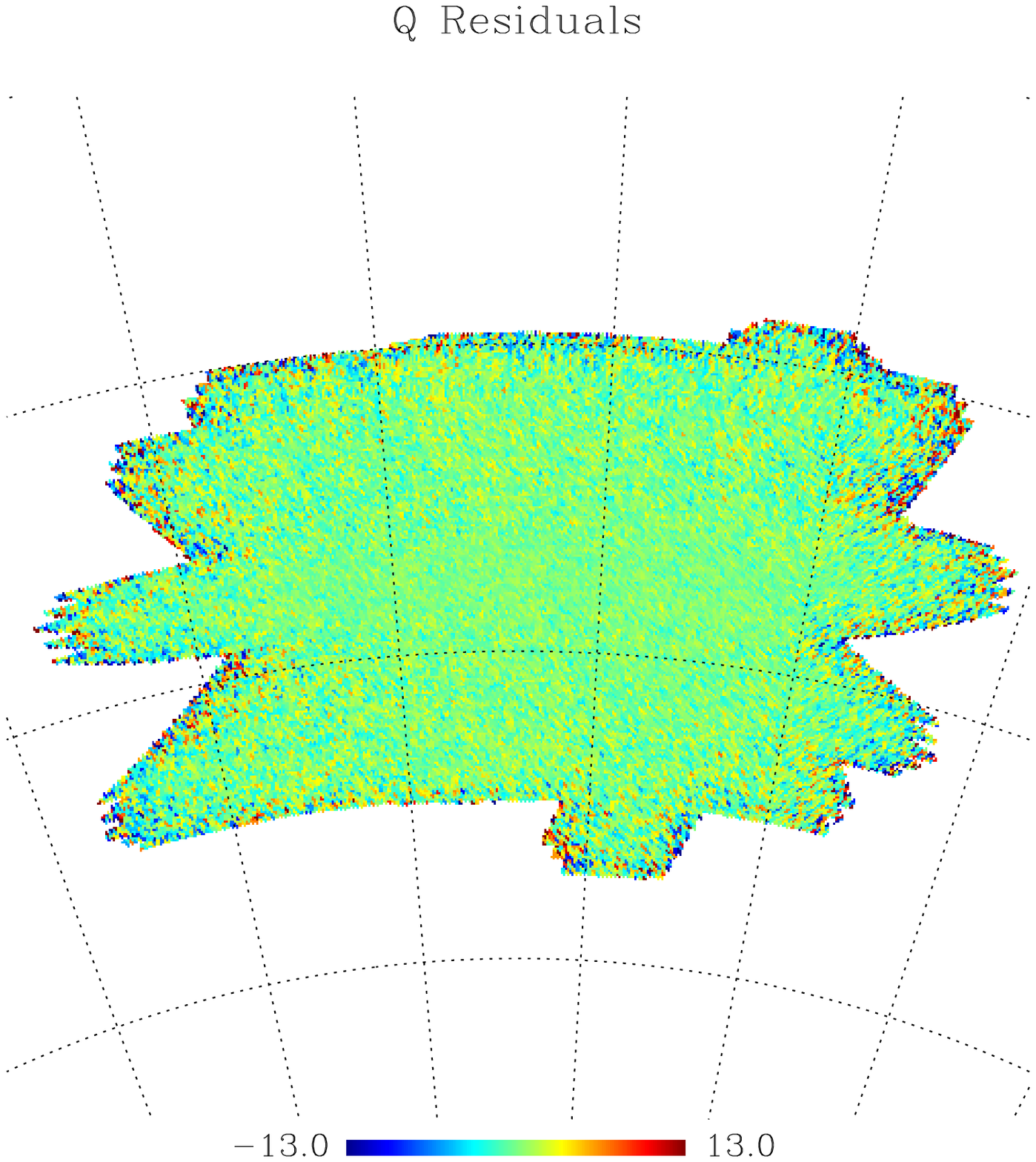}
\includegraphics[angle=90,width=0.4 \textwidth]{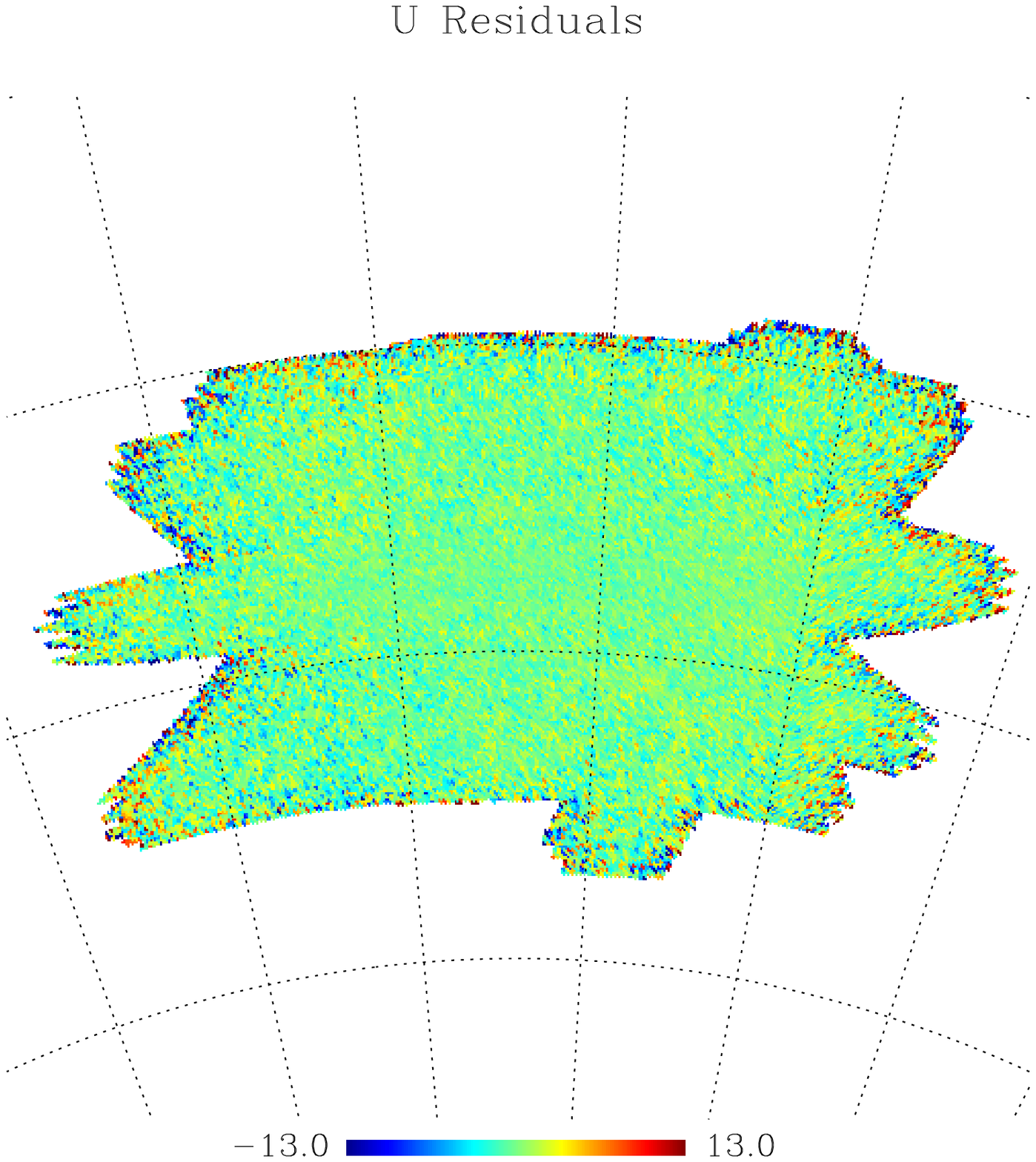}
\caption{Example maps of Q and U Stokes parameters in $\mu \textrm{K}$ from \textsc{descart} from simulations with white noise for 91 horns ($f_{k}= 200$-mHz and $\alpha=1.0$).  The top row is the map estimate and the bottom row is the residual map. A $10^{\circ}$ graticule is overlaid.}
\label{maps}
\end{center}
\end{figure*}

Figure \ref{maps} shows example Q and U maps from \textsc{descart} and their residual maps (defined as estimated map minus the map used to simulate the data).  The maps from the filtering pipeline are not shown, as they appear identical by eye. (For well cross-linked strategies, $1/f$ is a subtle systematic that appears most strongly in power spectra).  The comparative power in the filtered residual maps is dependent on the signal to noise ratio.  Filtering forcibly produces white noise in the output maps, optimally suppressing correlated noise.  In noise dominated maps, this makes filtering more effective in minimising map residuals.  

We compare the residual maps, which give the best measure of errors on the sky, by comparing the residual root mean square (rms) over the realisation ensemble.  Table \ref{rms residuals table} shows the rms residual for both pipelines for all the parameters simulated.  When the signal to noise ratio is small (19 and 91 horns), the residuals from filtering are smaller than those for destriping.  For the higher signal to noise simulations (1000 horns), destriping produces smaller residuals than filtering.  By a quirk of the parameterization, $1/f^{2}$ noise produces less noise at frequencies higher than $f_{k}$ than does $1/f$ noise, leading to larger residuals in noise case 3 compared to noise case 4.

\begin{figure*}
\begin{center}
\includegraphics[angle=90,width=0.4 \textwidth]{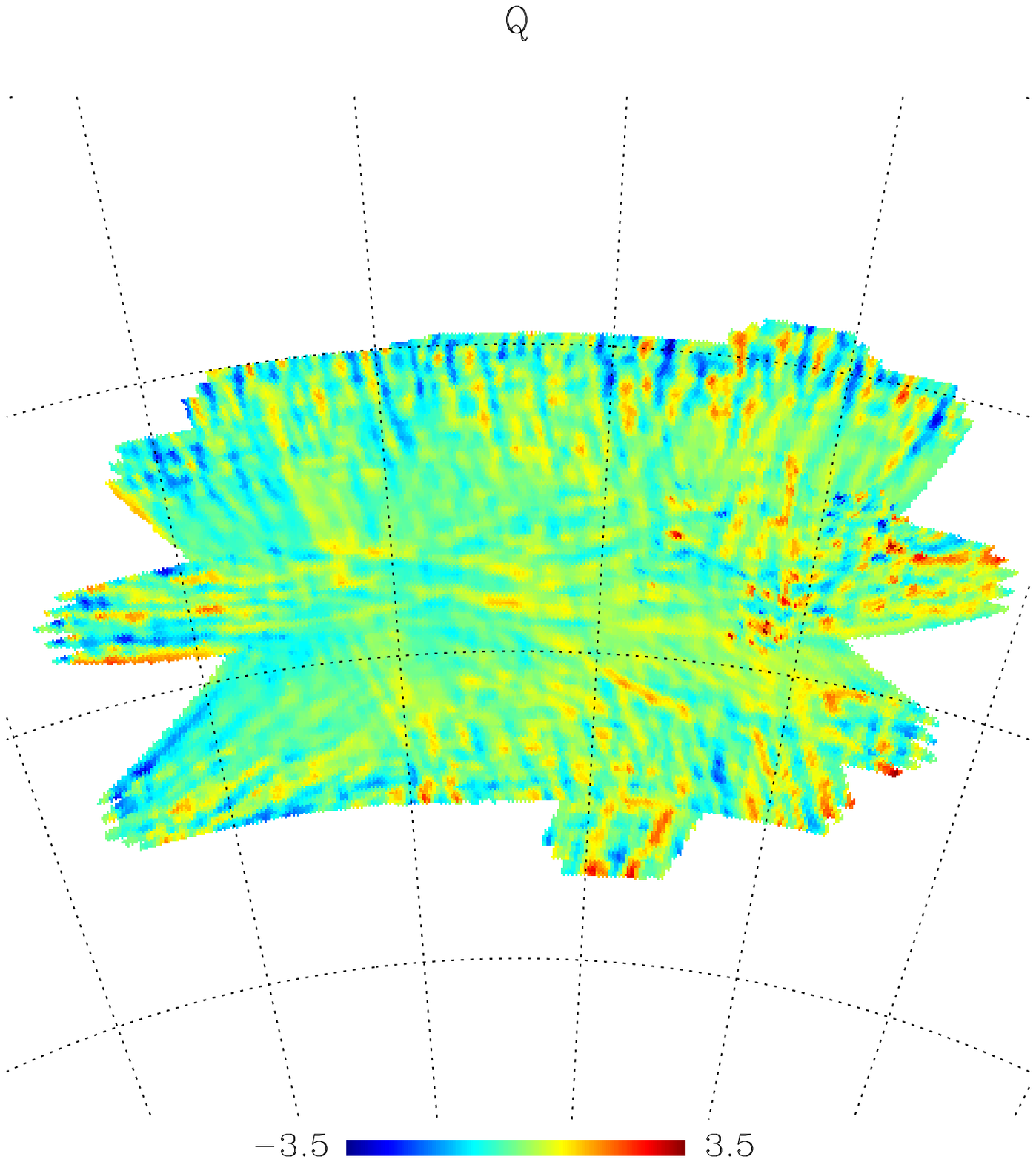} 
\includegraphics[angle=90,width=0.4 \textwidth]{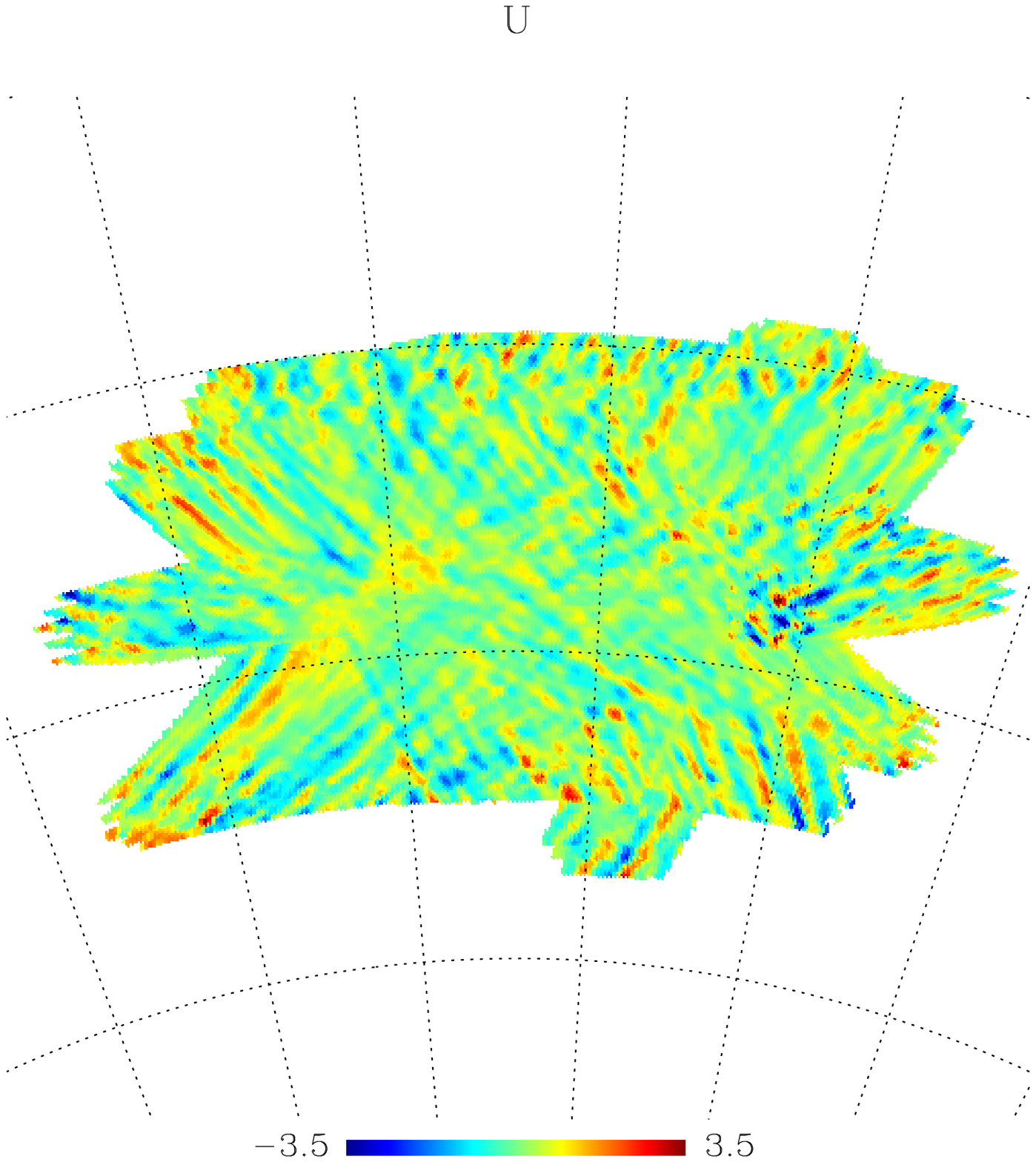}
\caption{Signal distortion in $\mu\mathrm{K}$ from (prewhitening) filtering in signal-only maps. The left panel is Q and right panel is U.}
\label{qu distortion maps}
\end{center}
\end{figure*}

Filtering suppresses signal as well as noise and in high signal to noise regimes, the signal distortion contributes appreciably to the residuals.  Signal distortion contributes more to the residuals when the required filtering is more aggressive, when either the knee frequency or spectral index of the correlated noise is increased.  Example maps of signal distortion are shown in Figure \ref{qu distortion maps}.  The magnitude of the signal distortion is not dependent on the integration time over a pixel, but rather on the level of cross-linking and polarisation angle dispersion in the pixel.

\begin{table}
\caption{RMS map residuals (in $\mu \textrm{K}$).   See Table \ref{parameters} for a description of the noise cases. The filter considered is the prewhitening filter.  The destriping length was $\lambda_{\textrm{d}}=1$-s for all except the $f_{k}=5$-mHz simulations (noise case 6), where it was $\lambda_{\textrm{d}} = 20$-s.}
\begin{center}
\begin{tabular}{ccccc}
noise case & descart Q & descart U & filtered& filtered U\\
\hline
1& 7.46&7.41&6.91&6.86 \\
2& 3.40&3.38&3.21&3.20 \\
3& 1.03&1.02&1.17& 1.19\\
4& 1.01 &1.00&1.18&1.20\\
5& 0.98 &0.97&1.04&1.04 \\
6& 0.94 & 0.93 & 0.99 & 0.99\\
\hline
\end{tabular}
\end{center}
\label{rms residuals table}
\end{table}

\subsection{B-mode systematics from filtering} \label{systematics section}

\begin{figure*}
\begin{center}
\includegraphics[angle=0,width=0.49 \textwidth]{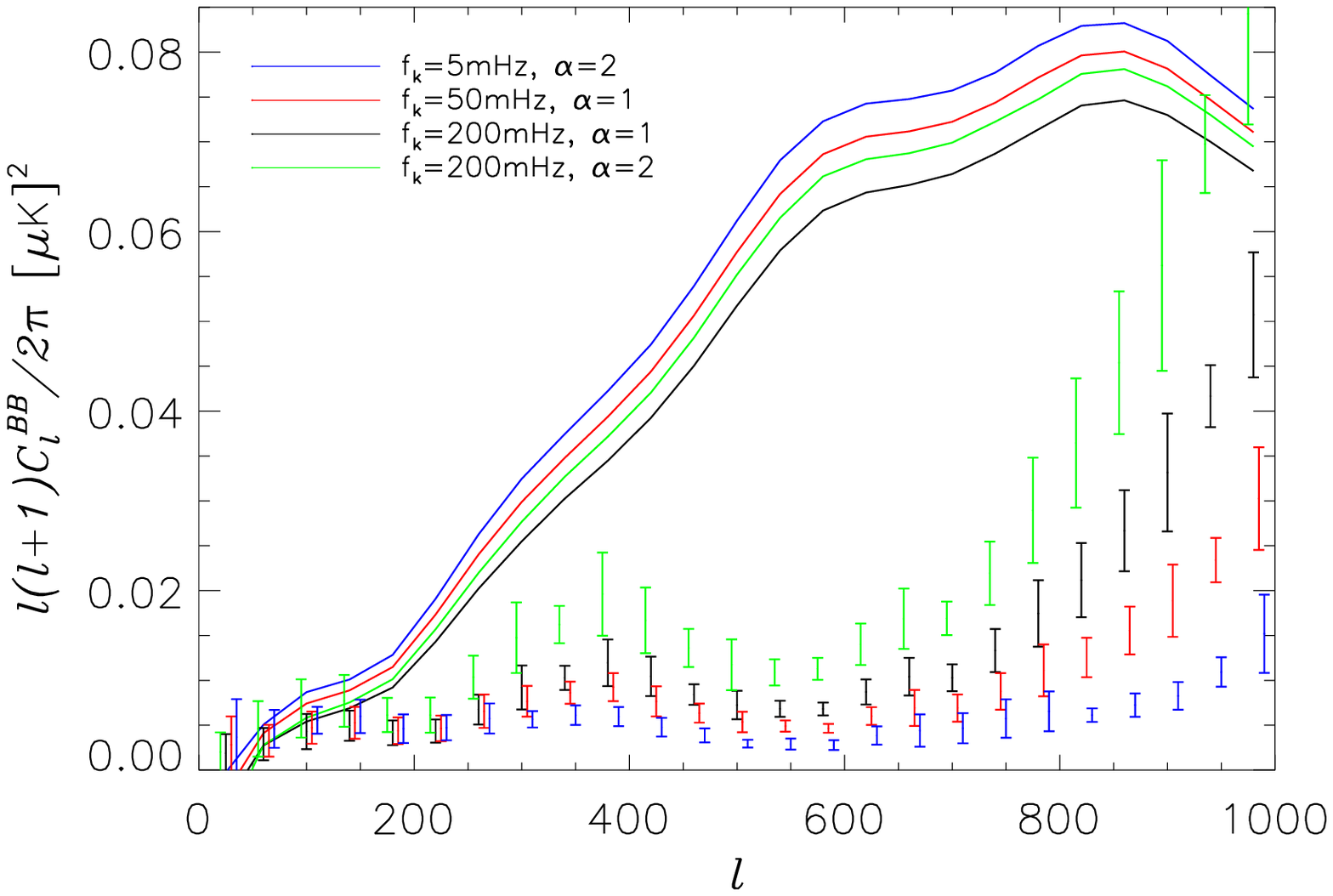} 
\includegraphics[angle=0,width=0.49 \textwidth]{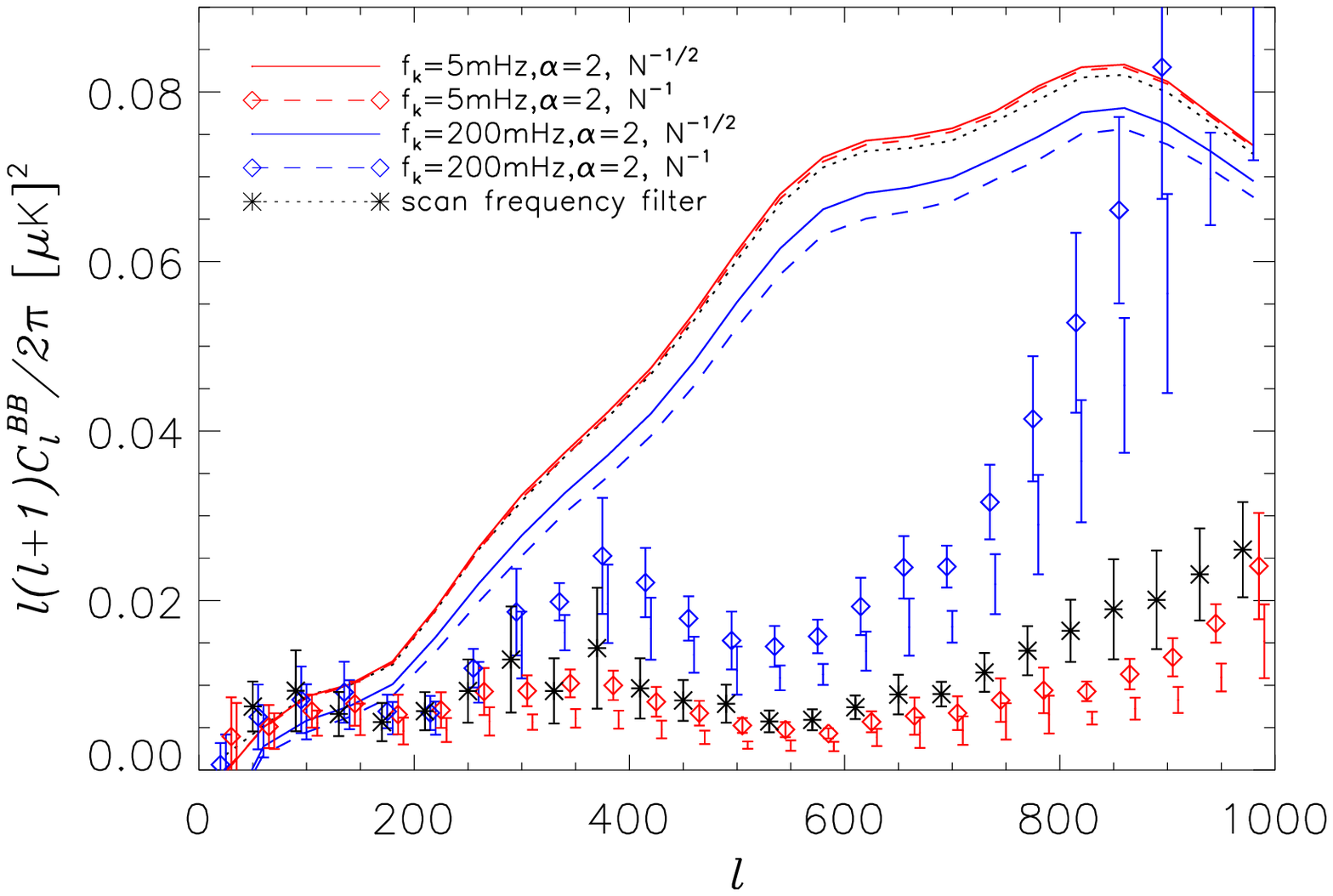} 
\caption{E$\rightarrow$B leakage systematic induced by filtering.  The left panel shows spurious B-mode power (error-bars) compared to the model B-mode power spectra (solid curves) suppressed by the prewhitening filter transfer functions.  The least aggressive filter (red curves) suppresses the input B-mode power spectrum the least and also distorts the Q and U maps the least, resulting in less E$\rightarrow$B leakage and less spurious B-mode power.  The red plots are for filtering with $f_{k}= 50$-mHz and spectral index $\alpha=1$, the black plots are for $f_{k}=200$-mHz and $\alpha=1$, the blue plots are for $f_{k}=5$-mHz and $\alpha=2$,and the green plots are for $f_{k}=200$-mHz and $\alpha=2$.  The right panel shows these effects for the overwhitening and scan frequency filters.  Colour denotes the noise case (blue is case 4 and red is case 6), with the solid curves and plain error bars showing the prewhitening filter, dashed curve and diamond error bars showing the overwhitening filter, and the dotted black curve and star error bars showing the scan frequency filter.}
\label{eb leakage figure}
\end{center}
\end{figure*}

The systematic error effects of TOD filtering are well understood in the case of temperature data (eg: \citealt{hivon:2002}), the dominant effect being the convolution of the estimated power spectrum with a filter transfer function, $F_{\ell}^{T}$. This can be determined from MC signal-only realisations and can be deconvolved from the estimated power spectrum.  In the case of polarisation, E and B-mode filter transfer functions can also be measured through simulations.

Our simulations show that B-modes estimated using the filtering algorithm are biased by a spurious B-mode component caused by E$\rightarrow$B leakage.  This leakage is absent when no filtering is applied to the TOD, as is the case with the destriping algorithm. It is also absent when the input maps are contrived to have no E-modes.

Figure \ref{eb leakage figure} shows the spurious B-mode from signal-only TOD simulations with zero input B-mode.  The left plot shows contrasting effects of TOD filtering: suppression of the observable B-mode power (by convolution with the filter transfer function $F_{\ell}^{B}$); and spurious B-mode power caused by distortions of the Q and U maps (such as those seen in Figure \ref{qu distortion maps}).   Both the B-mode suppression and the spurious B-modes are greater for more aggressive filters with higher knee frequencies and spectral indices.  The spurious B-mode is non-negligible for all of the noise cases considered at the band-powers of interest for measuring the primordial B-mode (which peaks at $\ell \sim 100$), with an amplitude of the order of the observable signal for $r=0.1$.

The right hand plot of Figure \ref{eb leakage figure} shows the spurious B-mode and signal suppression from overwhitening and scan-frequency filters in comparison to the prewhitening filter.  The overwhitening filters produce more spurious signal (and to a lesser degree, more signal suppression) than the equivalent prewhitening filter for the same noise case.  The scan frequency filter produces significant spurious signal, particularly at large angular scales, where it is larger than the observable B-mode signal.

\begin{figure*}
\begin{center}
\includegraphics[angle=0,width=0.49 \textwidth]{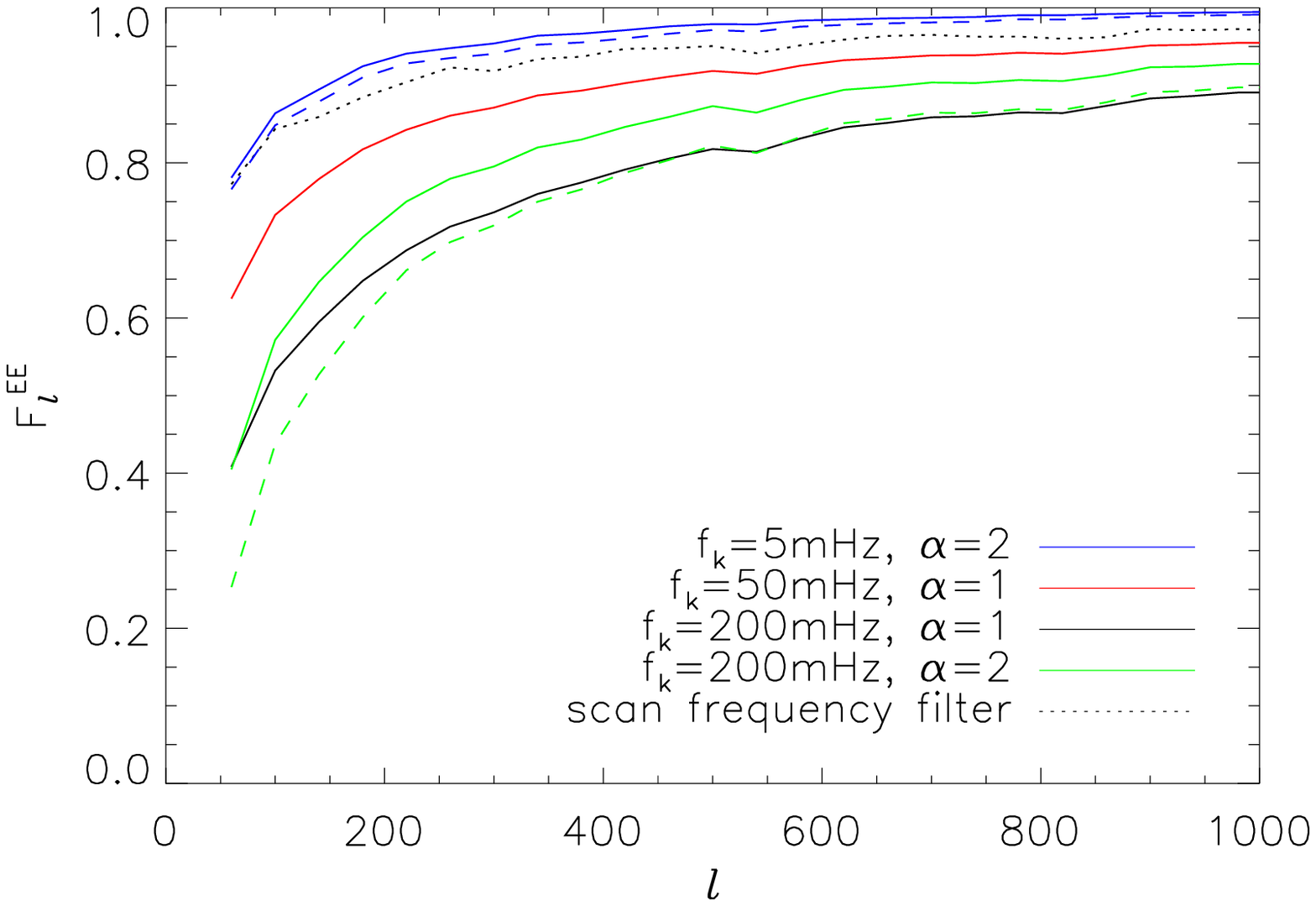}
\includegraphics[angle=0,width=0.49 \textwidth]{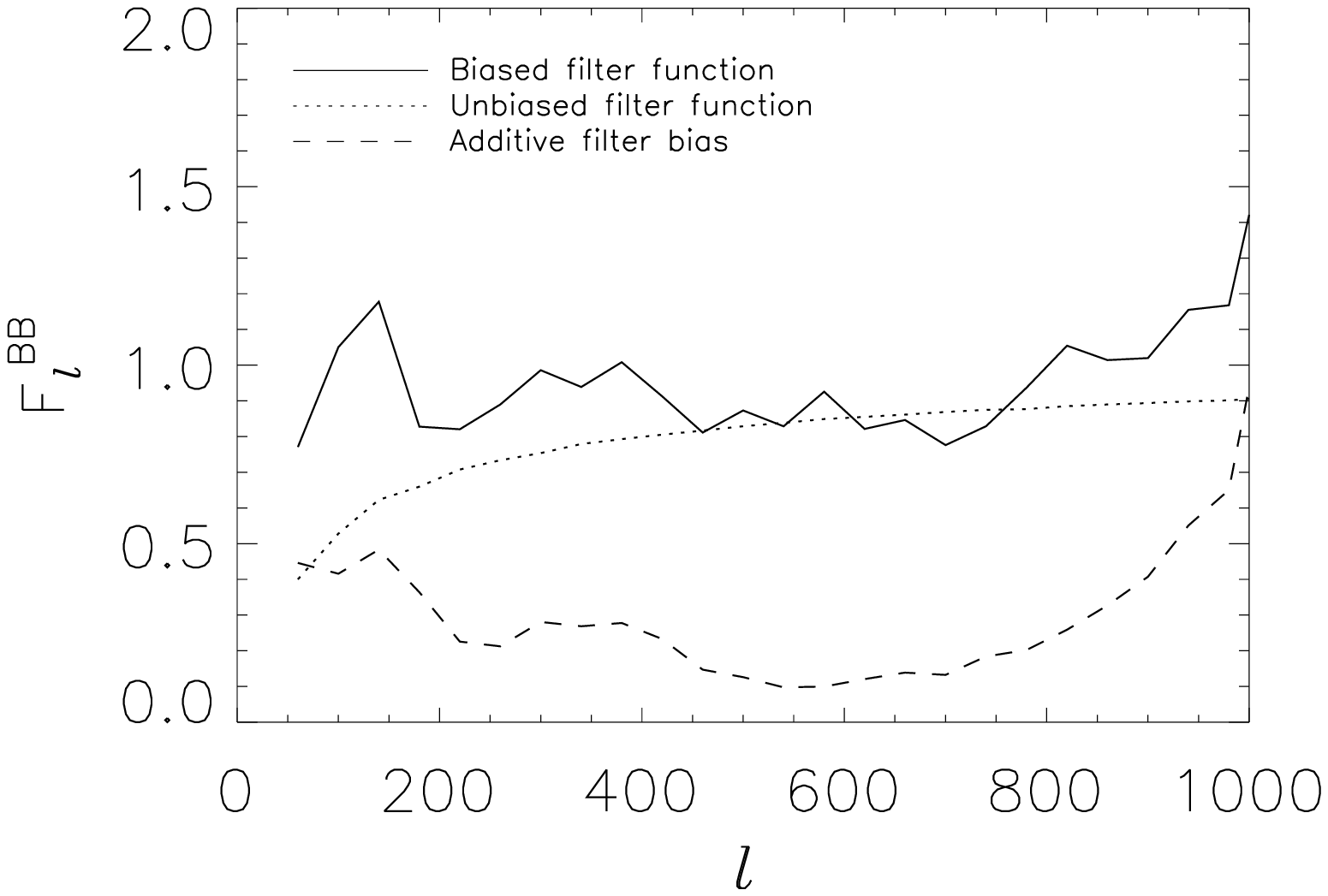}
\caption{Effects of noise correlation parameters on E and B-mode filter transfer function. The left plot shows the E-mode function for the filters considered.  The solid curves denote prewhitening filters and the dashed curves denote overwhitening filters, with the colour of both denoting the noise case.  We also show the filter transfer function for the scan-frequency filter.  The right plot shows the bias effect of the spurous B-modes on the filter transfer function $F_{\ell}^{BB}$ for the case with $f_{k}=200$mHz and $\alpha=1$.  The dotted curve is the unbiased filter transfer function $F_{\ell}^{B only}$ from B-mode only simulations, the solid curve is the biased B-mode transfer function $F_{\ell}^{B}$ and the dashed curve is the additive filter bias $F_{\ell}^{B+}$.}
\label{e_filter_fns}
\end{center}
\end{figure*}

The right hand plot of Figure \ref{e_filter_fns} shows the biasing effect of the spurious B-modes on $F_{\ell}^{B}$ as measured from MC signal-only simulations.  $F_{\ell}^{B}$ is the ratio of the $C_{\ell}$s of the filtered signal only map (including signal bias) and a binned map of signal only TOD
\begin{equation}
F_{\ell}^{B}
= \frac{\langle C^{B only}_{\ell} +  S_{\ell}^{B} \rangle}{\langle {C^{B}_{\ell}}^{bin} \rangle}
= F_{\ell}^{B only} + F_{\ell}^{+}
\end{equation}
where $S_{\ell}^{B}$ is the spurious B-mode signal that leads to an additive filter bias, $F_{\ell}^{+}$, of the filter function. The true B-mode filter transfer function is shown in the right panel of Figure \ref{e_filter_fns} (dotted curve, estimated from B-mode only signal simulations), compared to the additive filter bias (dashed curve, estimated from filtered simulations with zero input B-mode) and the resulting biased function (solid curve).  

The power spectra can be debiased by extending the \textsc{master} approach to include a spurious signal bias in the same way as it includes a noise bias.  If $K_{bb'}$ is the band-power coupling kernel as defined for polarisation in \cite{hansen:2003}, the unbiased power spectrum $\hat{C_{\ell}}$ can be returned by
\begin{equation}
\langle \hat{C_{b}} \rangle = K_{bb'}^{-1} P_{b'\ell}(\tilde{C_{\ell}} - \langle \tilde{N}_{\ell}^{MC}\rangle  - \langle \tilde{S}_{\ell}^{MC} \rangle)
\label{modified master estimator}
\end{equation}
where $P_{b\ell}$ is the binning operator defined in \cite{hivon:2002} and $N_{\ell}$ and $S_{\ell}$ are noise and spurious signal biases calculated over sets of MC simulations.

The $E \rightarrow B$ leakage contributes more than just power .  The variance of the leaked E-modes contributes to the variance of the debiased B-mode estimates (see \S \ref{power spectra errors section}).

The results presented in this section are limited to the single experimental set-up described in \S\ref{simulations section}.  We have found that the bias and its variance can vary with the details of the scan-strategy and focal plane arrangement, and consideration should be given to minimising this effect when implementing filtering methods.

\subsection{Polarisation power spectra errors} \label{power spectra errors section}

\begin{figure*}
\begin{center}
\includegraphics[angle=0,width=0.49 \textwidth]{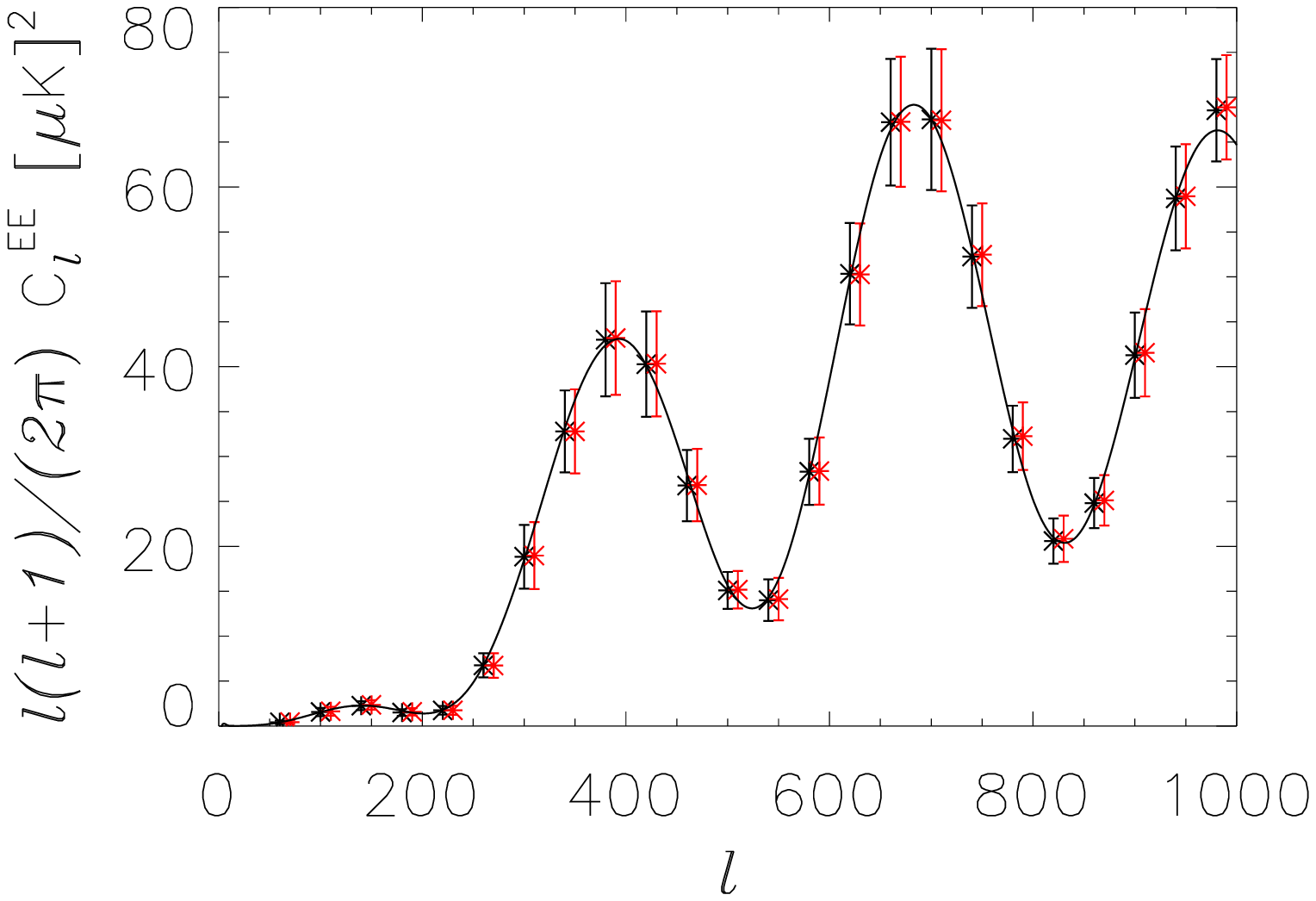}
\includegraphics[angle=0,width=0.49 \textwidth]{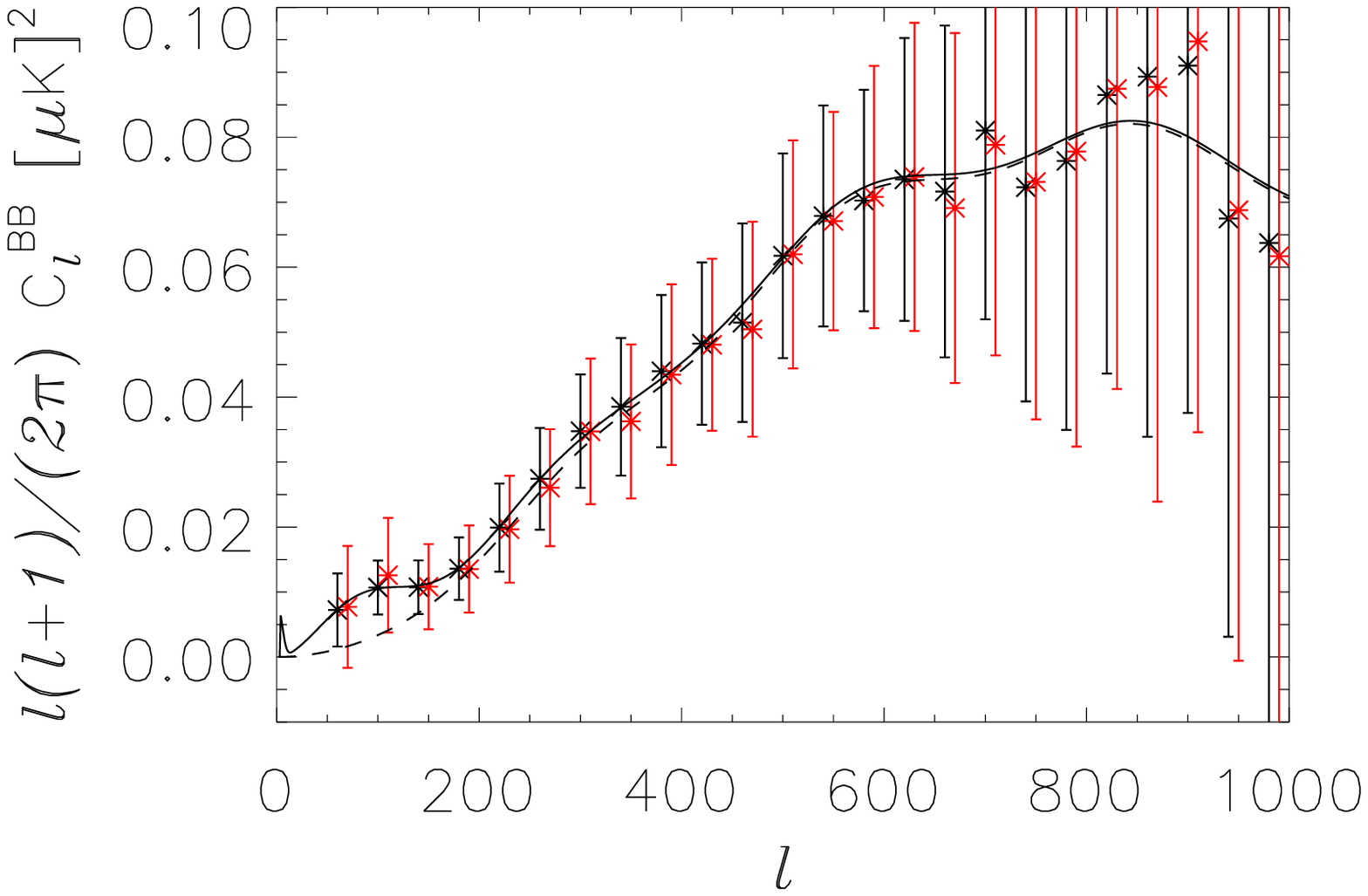}
\caption{Mean E-mode (left) and B-mode (right) estimates for simulations with $f_{k}=200$mHz and $\alpha=1.0$, with white noise for 1000 horns.  The black crosses are destriped estimates and the red crosses are filtered estimates.  The solid lines are the input fiducial power spectra and the dashed line in the B-mode plot is the lensing B-mode component of the input spectrum.  The large-scale B-mode bandpower variances are considerably enhanced for filtering compared to destriping, in this case preventing detection of the primordial B-mode peak.}
\label{EB estimates example plot}
\end{center}
\end{figure*}

Using both pipelines, we have estimated unbiased mean E and B-mode signals for each of the noise cases in Table \ref{parameters}.  In the filtering pipeline, the modified estimator (\ref{modified master estimator}) was used to remove spurious power from $E \rightarrow B$ leakage .  Figure \ref{EB estimates example plot} shows an example of the mean E and B-mode estimates over the ensemble of realisations for one of the parameter sets together with bandpower variances calculated over the realisation ensemble.  In this section, we evaluate the pipelines based on the bandpower variances of the debiased E and B-modes.

TOD filtering decreases the sensitivity of the power spectra to signal, particularly at large angular scales, as described by the filter transfer functions, $F_{\ell}$, shown in Figure \ref{e_filter_fns} for the different noise cases and filters.  These must be deconvolved from the power spectra estimates to produce an unbiased estimator of the CMB. 
Part of the gain of filtering noise correlations before mapping is counteracted by this loss of sensitivity to signal, as the signal-to-noise in heavily filtered modes is larger than can be achieved by optimal methods.  In our simulations, destriping consistently produces smaller variance in the power spectra than filtering.  Destriped power spectra do not require an $F_{\ell}$ deconvolution, as the estimator is by construction unbiased with respect to the sky signal and produces flat filter transfer fuctions. 

\begin{figure}
\begin{center}
\includegraphics[angle=0,width=0.49 \textwidth]{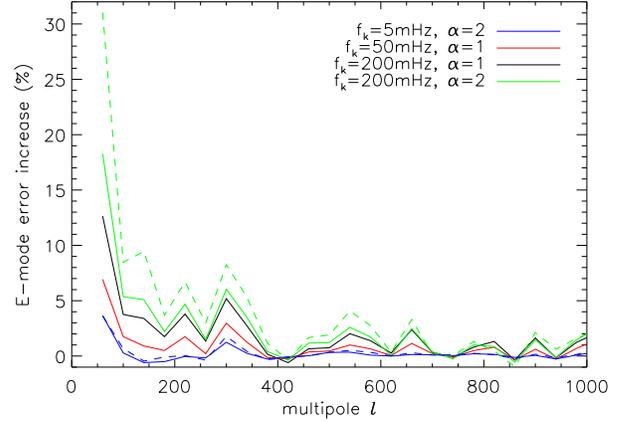}
\caption{E-mode error increase when using filtering instead of destriping, for simulations with varying $f_{k}$ and $\alpha$. The solid curves are for the prewhitening ($\mathbf{C_{N}^{-1/2}}$) filter and the dashed curves are for the $\mathbf{C_{N}^{-1}}$ filter.}
\label{e_err_incr}
\end{center}
\end{figure}

Figure \ref{e_err_incr} shows the ratio of the E-mode bandpower standard deviation from filtering to that from from destriping, presented as a percentage error increase, and its variation with the noise correlation parameters, $f_{k}$ and $\alpha$, for both prewhitening and overwhitening filters.  The error increase is larger for the more aggressive filters with higher knee frequency (or alternatively, filtering for simulations with more correlated noise).  Likewise, the more aggressive overwhitening filter results in larger bandpower variances at large angular scales than the corresponding prewhitening filter for the same noise case.

\begin{figure}
\begin{center}
\includegraphics[angle=0,width=0.49 \textwidth]{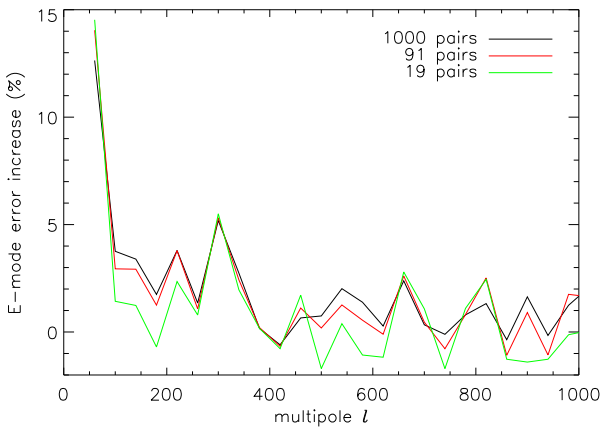}
\caption{Effect of signal to noise on E-mode error increase. For simulations with white noise level for 19, 91 and 1000 detector horns, using $f_{k}=200$-mHz and $\alpha=1$. Only $\mathbf{N^{-1/2}}$ filtering is considered.}
\label{e_err_incr_sn}
\end{center}
\end{figure}

The error increase is present in all signal to noise regimes simulated and shows only weak dependence on the signal to noise ratio in the maps.  Figure \ref{e_err_incr_sn} shows the effect of varying signal to noise on the E-mode error increase (for noise cases 1-3, with $f_{k}=200$-mHz and $\alpha=1$).    This contrasts with the map pixel variance, where filtering and destriping outperform one another in noise and signal dominated regimes respectively  (see Table \ref{rms residuals table}).  As the filter transfer function is scalar, it is not dependent on the magnitude of the white noise, but on the noise correlation statistics.  It does not depend on the ratio of the signal degradation in the map to the noise in the map and so the effect of white noise variation on the power spectra should be small.

\begin{figure}
\begin{center}
\includegraphics[angle=0,width=0.49 \textwidth]{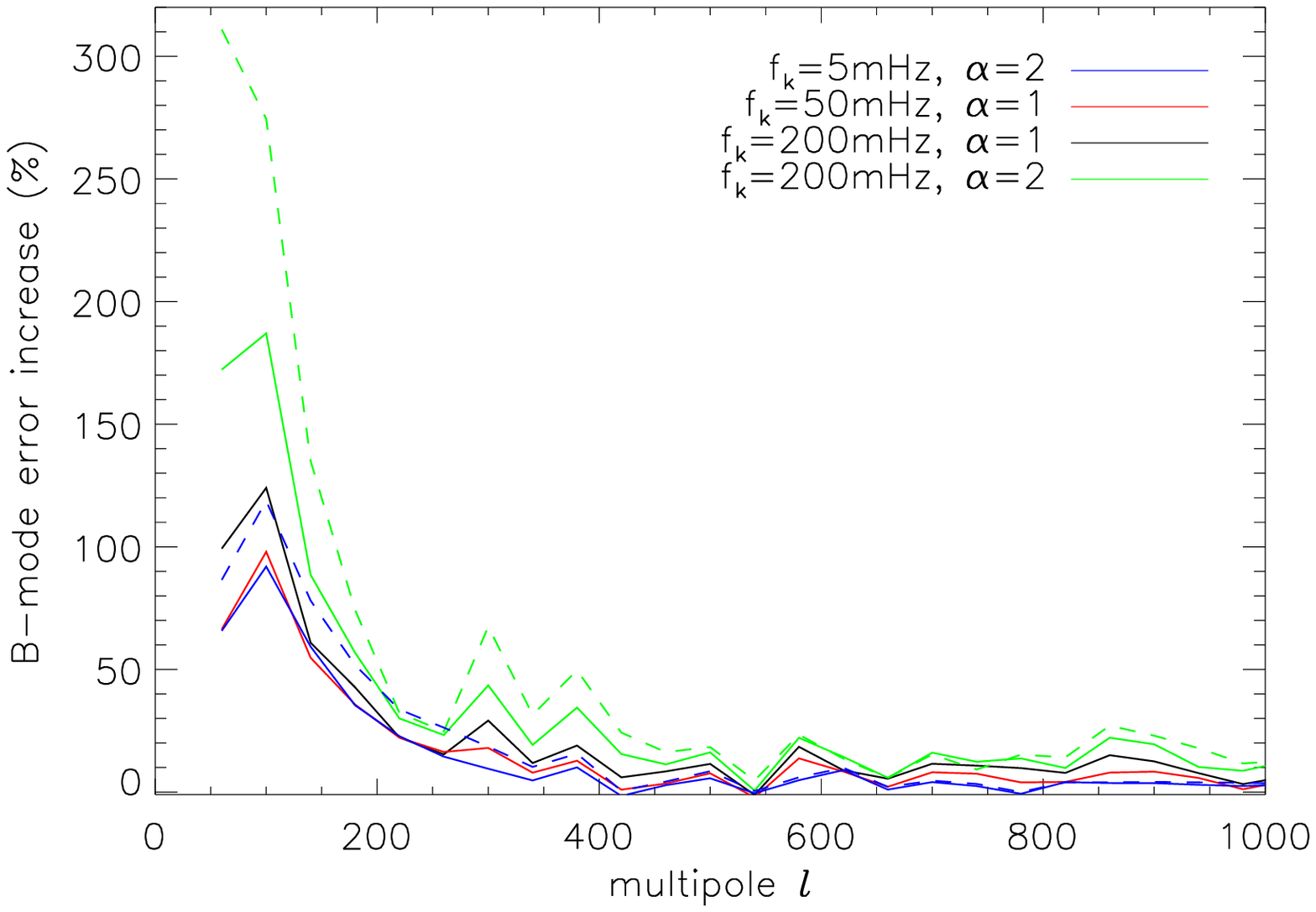}
\caption{B-mode error increase when using filtering instead of destriping, for simulations with varying $f_{k}$ and $\alpha$.  The solid curves are for the prewhitening ($\mathbf{C_{N}^{-1/2}}$) filter and the dashed curves are for the $\mathbf{C_{N}^{-1}}$ filter.  The error increase is larger than for the E-mode due to the additional variance from E$\rightarrow$B mixing by TOD filtering.}
\label{b_err_incr}
\end{center}
\end{figure}

B-mode bandpower errors are affected by both the loss of sensitivity due to the TOD filter and the extra variance from $E \rightarrow B$ leakage described in Section \ref{systematics section}.  Figure \ref{b_err_incr} shows the B-mode error increase for the different noise cases and filters.  The increase is considerably higher than for the E-mode, amounting to $> 50\%$ error increase at large angular scales for all of the noise cases simulated.  Low frequency noise filtering has a more pronounced effect on the error increase for B-modes than for E-modes, resulting in larger error increases for higher knee frequencies, higher spectral indices (in contrast to the E-mode, as discussed above) and for the overwhitening filter.  The disparity between this and the E-mode can be explained by the greater influence of the $E \rightarrow B$ leakage, for which variations in spectral index have a stronger effect.  As noted in the previous section, the size of leakage, and thus the error
increase reported here, can vary depending on the experimental setup.

\subsection{B-mode detection significance}

The primary goals of near-future CMB polarisation experiments are the detection and characterisation of the lensing component of the B-mode power spectrum, and the potential detection of the primordial gravity wave component. To achieve this we will require high precision from our analysis methods and any extra error they contribute to the angular power spectrum will have an impact on these science goals.

\begin{figure*}
\begin{center}
\includegraphics[angle=0,width=0.48 \textwidth]{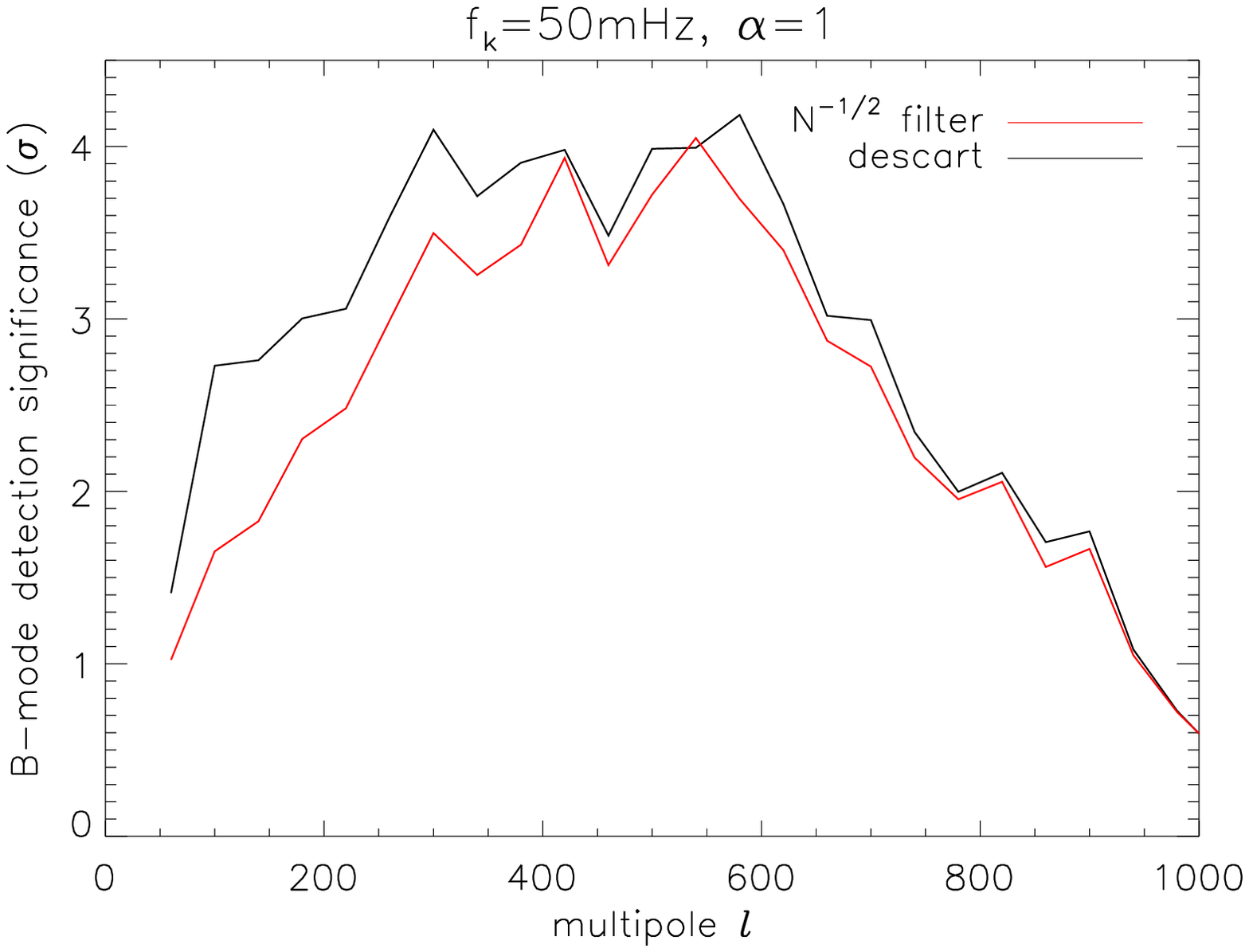}
\includegraphics[angle=0,width=0.48 \textwidth]{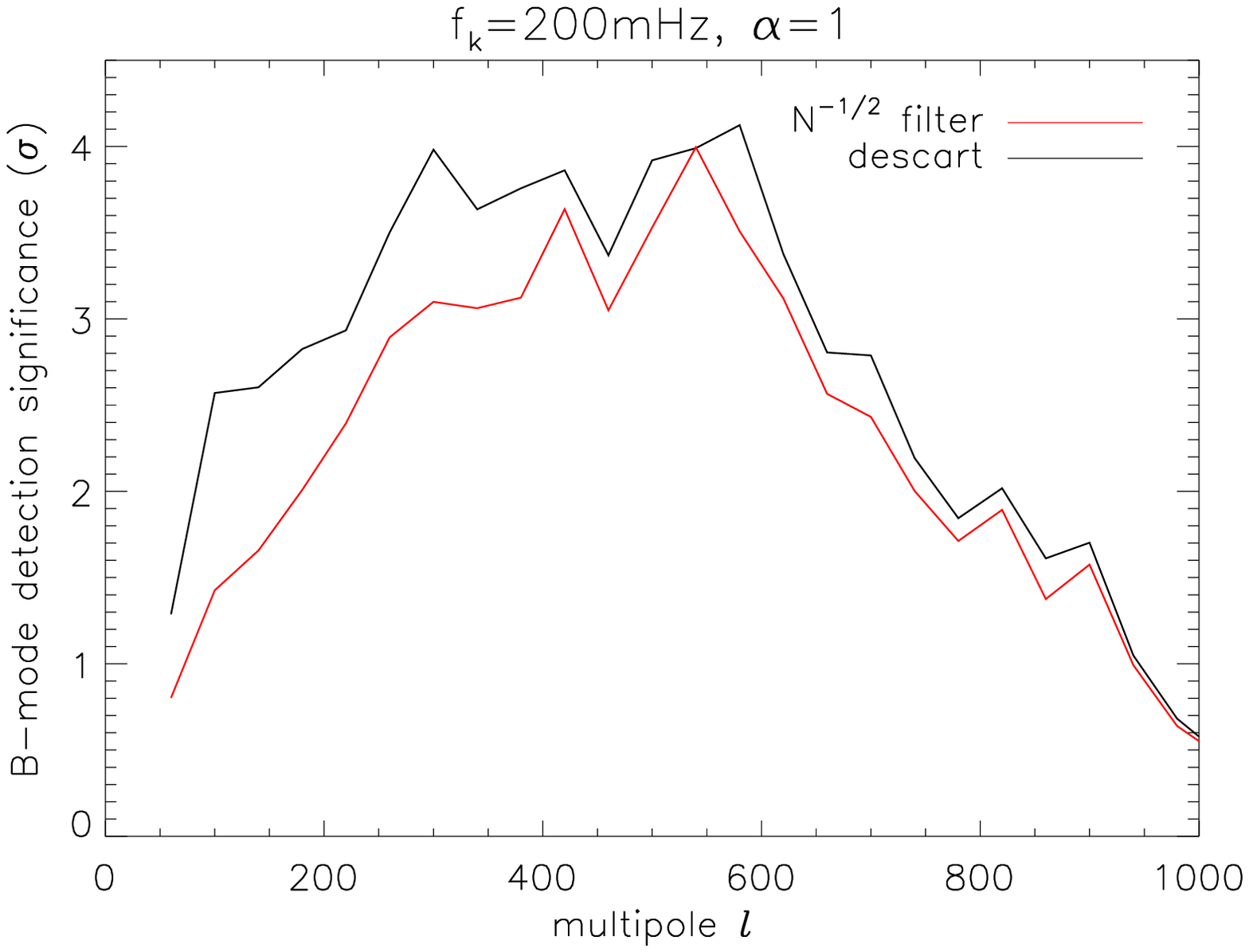}\\
\includegraphics[angle=0,width=0.48 \textwidth]{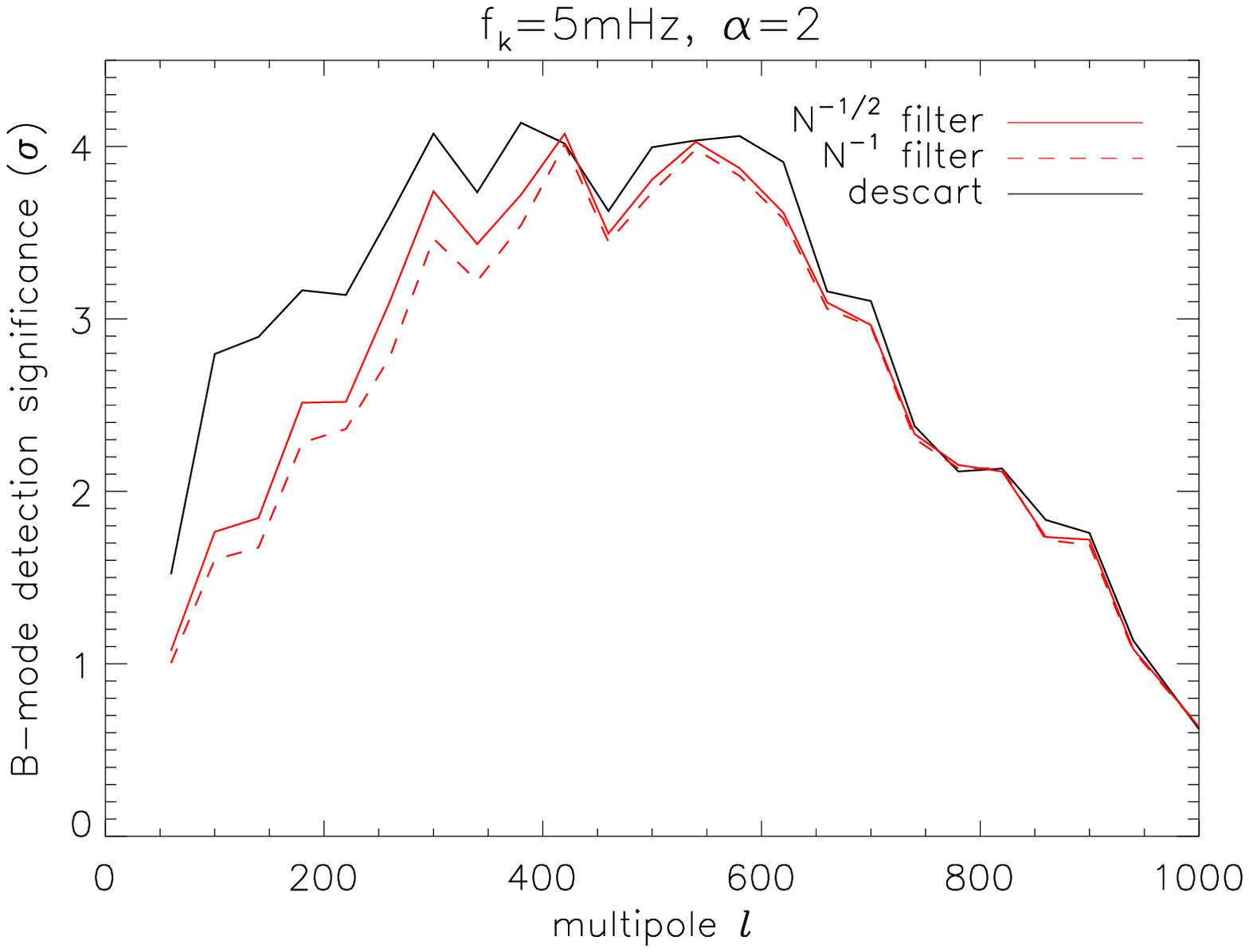} 
\includegraphics[angle=0,width=0.48 \textwidth]{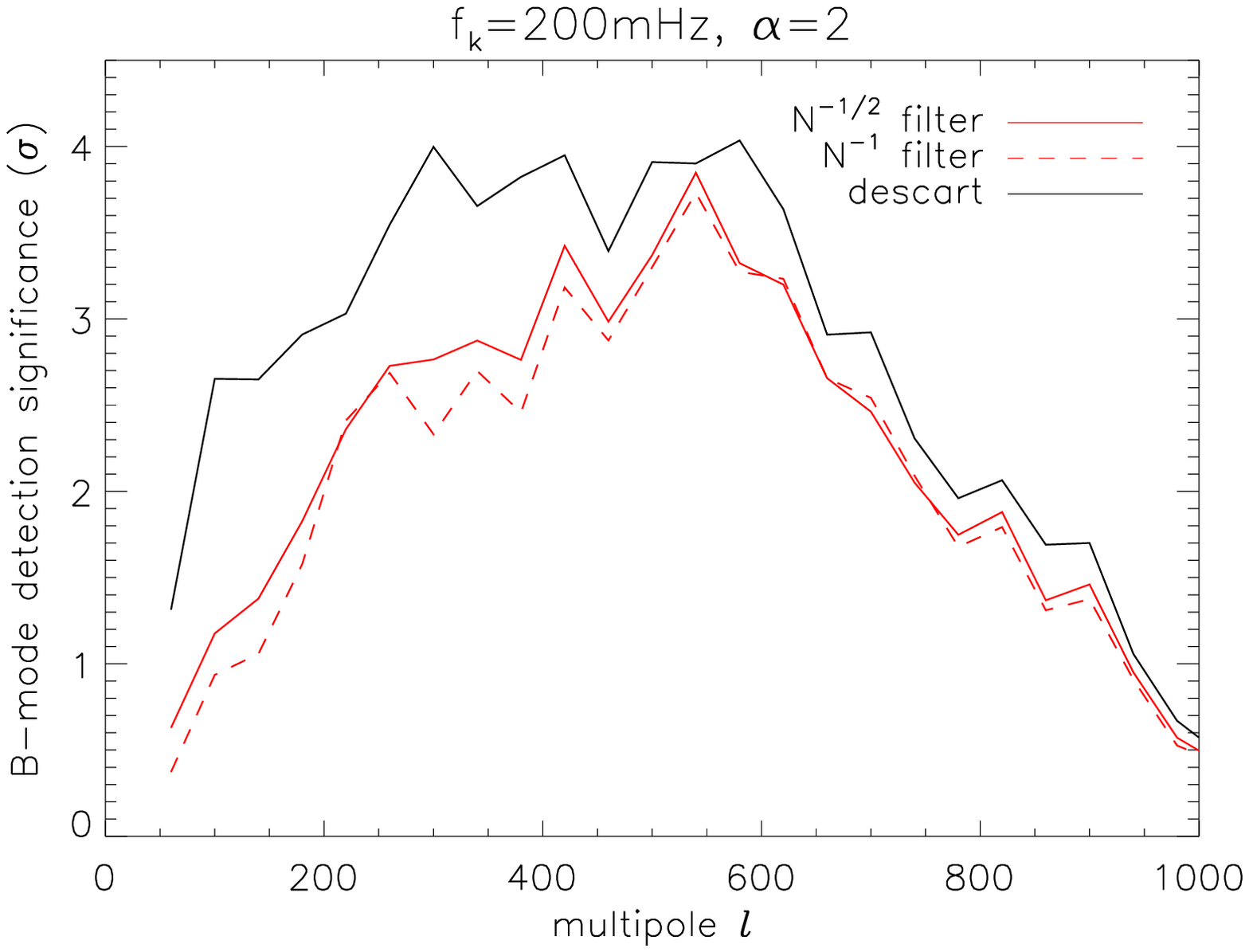}
\caption{B-mode bandpowers detection significance (in multiples of the bandpower error $\sigma$) for different noise correlation parameters, $f_{k}$ and $\alpha$, and signal to noise.  \emph{Top left}: $f_{k}=50$-mHz, $\alpha=1.0$ \emph{Top right}: $f_{k}=200$-mHz, $\alpha=1.0$. \emph{Bottom left}: $f_{k}=200$-mHz, $\alpha=2.0$. \emph{Bottom right}: $f_{k}=5$-mHz, $\alpha=2.0$.}
\label{significance plots}
\end{center}
\end{figure*}

To constrain the effect of bandpower variance differences on B-mode detection, we investigate the B-mode detection significance achieved by the pipelines.  Figure \ref{significance plots} shows the total B-mode detection significance per bandpower they return.  

The hump like shape of these curves is due to increasing bandpower variance due to Gaussian white noise at small angular scales and due to decreasing signal power at large angular scales.  Destriping produces higher significance bandpower detections, with the improvement increasing when the correlation of the noise (through $f_{k}$ and $\alpha$) is higher, or if the filter is more aggressive.

A directly comparable statistic can be calculated by combining the bandpowers into a single data point, measuring the total significance of the detection.  For this we use the total significance estimator defined in our previous paper \citep{sutton:2009}, that produces the single data point by binning weighted to a fiducial model $C_{\ell}^{fid}$ - in this case the known input B-mode power spectrum - and weighted inversely by the the bandpower variance $\sigma_{b}^{2}$
\begin{eqnarray}
\hat{C}= \frac{\sum_{b} (C_{b}^{fid}/\sigma_{b}^{2})\hat{C}_{b}}
{\sum_{b} (C_{b}^{fid})^{2}/\sigma_{b}^{2}}
\end{eqnarray}
Table \ref{total significance table} shows the B-mode detection significance, defined as $[\hat{C} / \sqrt{\langle (\hat{C} - \langle \hat{C} \rangle)^{2} \rangle}]$.

We also show the significance of the detection of the primordial B-mode peak in the bandpower centered at $\ell=100$.  Whilst our chosen input noise level has not supported a $2\sigma$ detection for the peak, it shows filtering can shave almost $1 \sigma$ off the weak detection from destriping.

\begin{table}
\caption{Total significance of primordial plus lensing B-mode detection.}
\begin{center}
\begin{tabular}{ccccc}
$f_{k}$ & $\alpha$ & descart & $C_{N}^{-1/2}$ filter & $C_{N}^{-1}$ filter \\
\hline
$5$-mHz & 1.0 & 13.44 & 12.49 &12.12\\
$50$-mHz &1.0& 13.05 & 11.56  & - \\
$200$-mHz &1.0& 12.64 & 10.69 & - \\
$200$-mHz &2.0& 11.77 & 9.98  & 9.44\\
\hline
\end{tabular}
\end{center}
\label{total significance table}
\end{table}

\begin{table}
\caption{Significance of B-mode inflationary peak detection.}
\begin{center}
\begin{tabular}{ccccc}
$f_{k}$ & $\alpha$ & descart & $C_{N}^{-1/2}$ filter & $C_{N}^{-1}$ filter\\
\hline
$5$-mHz & 1.0 & 1.87 & 1.28 & 1.18 \\
$50$-mHz &1.0& 1.83 & 1.20 & - \\
$200$-mHz &1.0& 1.72 & 1.05 & - \\
$200$-mHz &2.0&  1.78 & 0.87 & 0.70 \\
\hline
\end{tabular}
\end{center}
\label{bmode peak table}
\end{table}

\section{Application to a massively multi-detector experiment} \label{quiet simulations section}
The speed of map-making methods will become vital in future CMB experiments, as data-set sizes increase.  It is important therefore to understand the algorithmic scalings of the methods available and whether destriping is competitive as a fast map-maker.  Destriping is linear with the number of pixels in the map, making it useful as a method for high resolution analysis, but its scaling with data-set size is the more relevant question for massively-multidetector small field experiments.

In this section, we address this question by applying the \textsc{descart} code to simulated full-season data from a massively multi-detector experiment to determine the scalability of the method to much larger datasets than those in the previous section.  In \S \ref{algorithmic scalings section}, we determine the scaling of computing time and iteration number with the dataset size.  In \S\ref{analysis partitioning section}, we investigate the effect on the map pixel variance of partitioning the analysis by scan, a procedure that may be necessary for very large datasets.

The simulations were produced using the pointing information from the first full-season of QUIET Q-band operation.  The simulated dataset comprises of 202 constant-elevation scans (CESs) of a CMB sky patch with a total wall-clock observation time of 704.6 hours at a sampling frequency of 50-Hz.  Each CES includes timestreams from 68 detetors - produced by 17 hexagonally arranged horns with four polarisation sensitive diodes per horn.  This produced $4.7*10^{4}$ hours of integration time, summed over all the detectors. The total memory requirement for the whole dataset was $67$-Gb, rising to 101-Gb including pointing information.

$1/f$ noise was simulated to be uncorrelated between detectors and between CESs.  We choose a noise $f_{k}$ of 200-mHz and an offset length of $\lambda_{\textrm{d}}=2$-s.  These simulations were produced to determine the scaling of software and use typical, fiducial noise characteristics.  As such, neither the noise characteristics nor the simulations they produced are similar to real QUIET data.

\subsection{Algorithmic scalings} \label{algorithmic scalings section}

We have investigated the scaling of computing time with data-set size by applying \textsc{descart} to an increasingly larger set of CESs.  In this scheme,  all of the included data is used to estimate and subtract the naive map of the offset solutions - the operation $\mathbf{ZF}\vec{a}$ from equation (\ref{amp_destr}).  Convergence of the PCG iterator is achieved when the offsets from all of the included time-streams satisfy the convergence critereon - that the error vector $\lVert \epsilon \rVert = \lVert Ax-b \rVert < \tau  \lVert b \rVert$, where $\lVert \rVert$ denotes the 2-norm and the arbitrary stop tolerance $\tau = 10^{-6}$.  Using all of the available data to estimate the offset vector $\vec{a}$ means that the size of the system to be solved becomes very large when realistic datasets are involved.

\begin{figure}
\begin{center}
\includegraphics[angle=0,width=0.48 \textwidth]{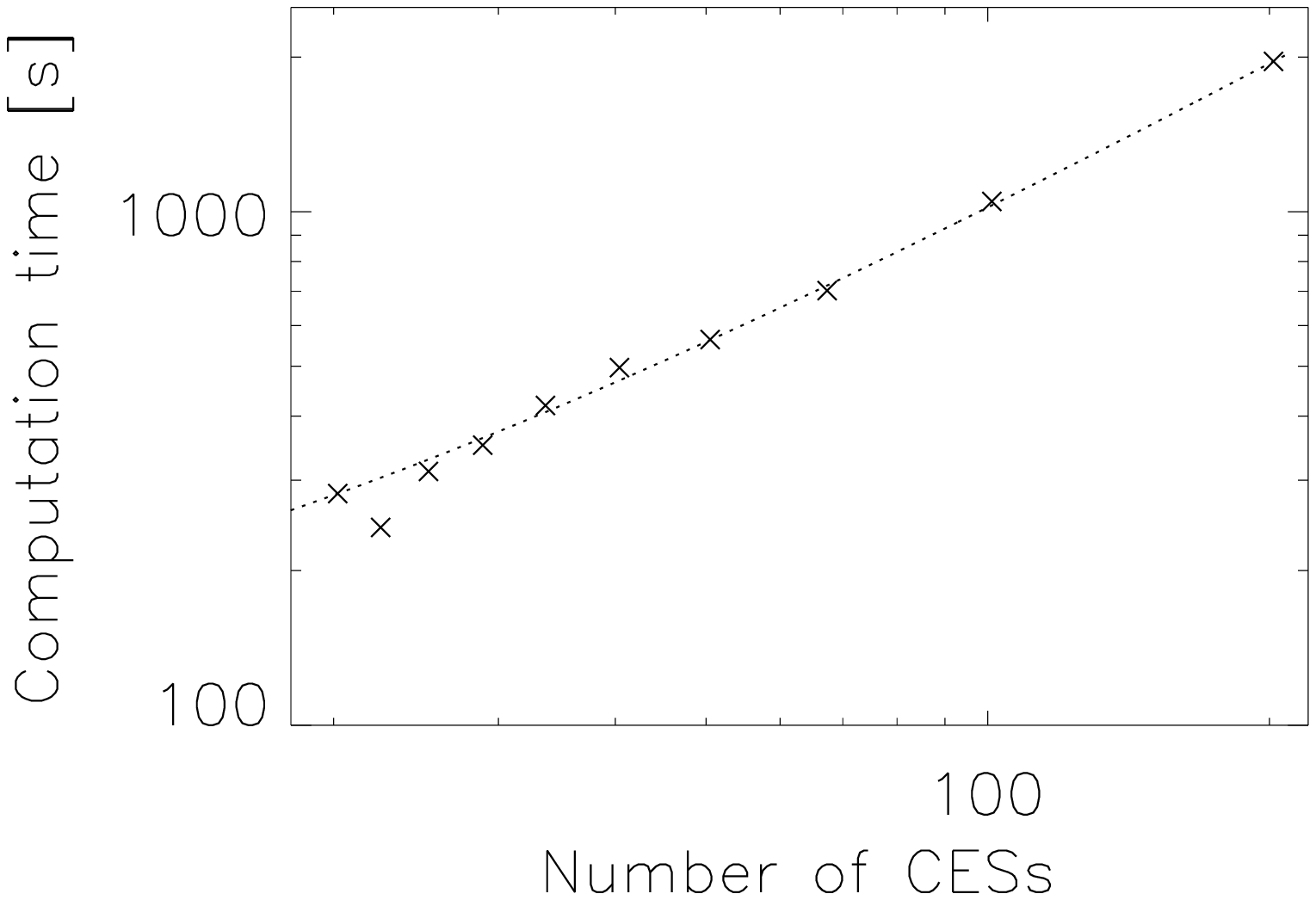}
\includegraphics[angle=0,width=0.48 \textwidth]{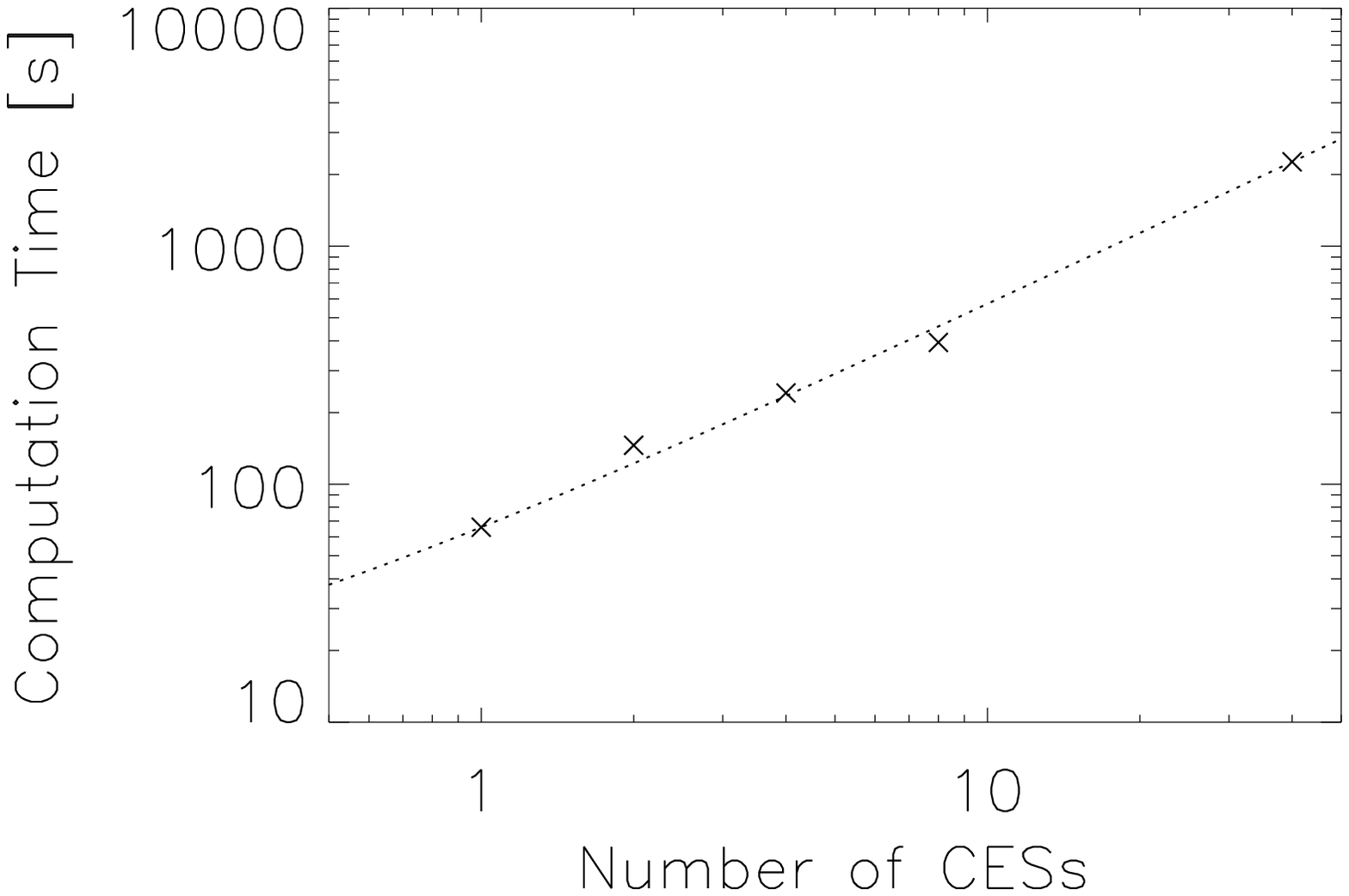}
\caption{Linear scaling of computation time with number of CESs included in destriping analysis, with a dotted linear curve shown for reference.  \emph{Upper panel}: Mean computing time per CES set for whole data-set (202 CESs) using 64 processors. \emph{Lower panel}: Mean computing time per CES set for smaller dataset (40 CESs) using 16 processors.} 
\label{computation time plot}
\end{center}
\end{figure}

However, the computation time of the algorithm scales linearly with the quantity of input data.  Figure \ref{computation time plot} shows the scaling of computing time with the number of CESs included in the analysis.  The computing time does not include the initial data-load operations, which will be system architecture dependent, and the time plotted is an average: the total computing time required to analyse all 202 CESs for the upper panel, and 40 CESs for the lower panel, divided by the number of partitions of that size required to analyse the whole data-set.  In our analyses of the maps, the majority of the total computation time was taken by the data load.

The scan sets shown in both panels of Figure \ref{computation time plot} were processed with \textsc{descart} on the 610 node cluster Titan, at the University of Oslo \footnote{http://hpc.uio.no/index.php/Main\_Page}.  The larger scan sets in the upper panel were processed using 64 cores on 8 nodes, whilst the smaller sets of the lower panel were processed using 16 cores on 2 nodes, so the panels are not directly comparable.

The operation count per iteration is dominated by the signal subtraction operation (\ref{definition of Z}) that includes two large linear $\mathcal{O}(N_{t})$ processes - map binning and map projection.  These dominate the smaller $\mathcal{O}(N_{\alpha} \log N_{\alpha})$ FFT operations associated with the prior $\mathbf{C_{a}}$, as $N_{\alpha}$ is two orders of magnitude smaller than $N_{t}$.  The total computation time depends on the number of iterations taken by the PCG algorithm to return a sufficiently precise estimate of the offset amplitudes.  For the scanning strategies we consider, the number of iterations does not increase as more data is added: rather it decreases.  This behaviour is shown in Table \ref{iterations table}, which shows the mean iteration number required per CES set.  This can be understood by noting that increasing the number of CESs included in the analysis provides a better estimated naive map in the $Z$ operator of (\ref{definition of Z}), so the operation $\mathbf{ZF}\vec{a}$ becomes closer to $\mathbf{F}\vec{a}$ (an effect enhanced by the lack of correlation in noise between CESs).  With less noise wrongly identified as signal, the condition number of the destriping matrix in (\ref{amp_destr}) decreases and so the number of iterations required to solve the simpler set of equations reduces.

The reduction in the iteration number is most marked when CESs are considered in pairs instead of independently.  This is due to the increased level of cross-linking in the combined scanning strategy of the two CESs, both in revisitations of the map pixels and in variation of the polarisation-sensitivity angles for the pixel.  The vast majority of the reduction in iteration number in our simulations is achievable by including groups of 4 CESs.

The above arguments can be applied to increases in the number of horns.  In the absence of inter-horn correlated noise, further horns are computationally identical to further CESs (other than in terms of cross-linking).  We therefore argue that the scaling of the algorithm with horn numbers will also be linear.

No inter-timestream cross-correlated noise was included in the simulations for this section, so the effect of including a cross-correlated prior as in \S\ref{destriping correlated streams section} was ignored.  Including this effect results in an additional operation per iteration, in which the inverse of the Fourier space inter-detector prior matrix is applied to the Fourier space data (equation \ref{mat mult}), which does not change the number of FFTs applied each iteration.  The operation scales as  $\mathcal{O}(n_{corr} N_{d}^{2} N_{\alpha})$, where $N_{\alpha}$, $N_{d}$ and $n_{corr}$ are the numbers of destriping offsets, detectors and correlated pairs for each CES.  In the case that the number of detectors and correlated detector pairs are high, this operation could come to dominate the $\mathcal{O}(N_{t})$ operations discussed above.

\begin{table}
\caption{Variation of mean iteration number with the number of CESs included in the analysis.  For CES numbers between 40 and 202, the number of iterations remains constant.  The mean iteration number is defined as the total number of PCG iterations required to map the whole dataset divided by the number of CES sets required to analyse it.}
\begin{center}
\begin{tabular}{cc}
\# CESs & \# Iterations \\
\hline
1 & 27.25 \\
2 & 17 \\
4 & 16.3 \\
8 & 15.4 \\
40 & 14 \\
\hline
\end{tabular}
\end{center}
\label{iterations table}
\end{table}%

\subsection{Analysis Partitioning} \label{analysis partitioning section}

The memory requirement for analysing complete data-sets becomes untenable for massively-multidetector ground-based experiments, so partitioning the data for the destriping step is an attractive analysis option.  Destriped maps can be combined with a noise-weighted average - an operation that is mathematically identical, other than through the accuracy of the offset estimates, to making a map from the entire destriped dataset.

However, the accuracy of the offset amplitude estimation can be increased by including more data in the solution of equation \ref{amp_destr}.  As argued in \S\ref{algorithmic scalings section}, the signal removal operator $\mathbf{Z}$, which removes both signal and any noise that looks like signal, removes less noise if more data is considered.  If it were to remove no noise at all, such that $\mathbf{Z}\vec{n} = \vec{n}$, then the offsets become the reference offsets described in \S\ref{timestreams and maps section}.  We can expect that the pixel noise in the destriped maps will reduce as the offset amplitude estimates better model the correlated noise in the timestreams.

Destriped maps were produced for different destriping partitions, then combined to produce a final map that uses all of the available data (in this case, the 40 CES data-set).  For a set of maps $\vec{x_{i}}$, with Q/U white noise covariances $\mathbf{C_{i}}$, the combined map is defined as $\vec{x} \equiv (\sum_{j} \mathbf{C_{j}}^{-1})^{-1} \sum_{i} \mathbf{C_{i}}^{-1} \vec{x_{i}}$.

\begin{figure}
\begin{center}
\includegraphics[angle=0,width=0.48 \textwidth]{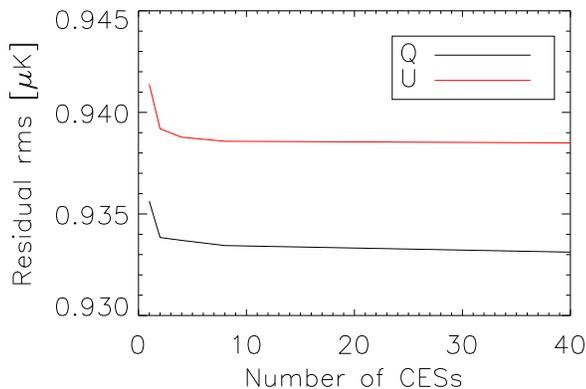}
\caption{Variation in RMS residual for the Q and U Stokes parameter maps with destriping partition size.  The maps were combined such that the same quantity of data was used for each map.  The variation seen is therefore entirely due to offset estimation error, as opposed to white noise level in the maps.}
\label{partitioning residuals figure}
\end{center}
\end{figure}

This ensures that the noise amplitude in the final maps is constant and any differences between them are solely due to changes in the error of the offset amplitude estimates.  Figure \ref{partitioning residuals figure} shows the variation of the RMS residual of Q and U in the final combined map with destriping partition size.  Only pixels within the central region of the scan (the science field itself) were used to calculate the RMS residual.

The effect of partition size on RMS residual amounts to a $0.3\%$ reduction between destriping with 40 CESs and destriping with single CESs only.  Of this change, the majority of the improvement is achieved by considering pairs of CESs - as was the case with iteration numbers and for the same reasons.  The residual RMS remains constant for CES numbers up to the full 202 CES dataset.

A number of effects that naturally produce cross-linking within a single CES contribute to make this change small.  In a separate analysis, we considered data from the central horn only, to remove the cross-linking effects of the 17 horn focal plane.   The variation in RMS with CES number is qualitatively similar to that shown in Figure \ref{partitioning residuals figure}, except that the RMS reduction is much larger - amounting to a $3\%$ reduction in residuals when all 40 CESs are used to estimates offset amplitudes rather than considering each CES separately.

The effect of destriping analysis partitioning has been investigated recently for the Planck experiment by \citep{kurki-suonio:2009}, who found a significant variation in destriping pixel residual RMS with the length of scan considered.  However, we note that the scanning strategies of the Planck and QUIET experiments have little in common - the Planck scanning strategy used revisited the vast majority of pixels only after a 6 month spin period, with the destriping relying on the cross-linking in a small number of polar pixels to determine amplitudes for offsets the length of a single great circle scan.  The scanning strategy considered here is heavily cross-linked for most pixels on small time-scales.

Our simulations suggest that destriping partitioning is an acceptable analysis technique when the partition consists of a few CESs.  A caveat for this result is that it is based solely on the effects of $1/f$ noise.  Instrumental systematics, the simulation of which is beyond the scope of this paper, can be better constrained with larger data partitions.  An example of this is scan-synchronous noise, which can be modelled by the offsets when large numbers of CESs with uncorrelated scan-synchonous noise are included, and which significantly increases the number of iterations required to solve for the offset amplitudes.

\subsection{Resources Summary}
A year of data consisting of 600 hours of data with 70 detectors 
sampled at 50Hz can be analyzed with this code on 64 typical modern 
processor cores in approximately 30 minutes.  This will scale 
approximately linearly with increasing time and number of detectors.

This is fast enough that destriping can be included as part of a Monte-
Carlo pipeline, replacing map-making with heavy filtering with only a 
small increase in computing time.  It is also suitable as an 
alternative to maximum-likelihood mapmaking for performing repeated 
auxiliary tasks such as null tests in cases where running the full 
pipeline repeatedly is unfeasible.

The code and method can therefore be applied to the largest presently 
existing CMB polarization data sets to make high resolution maps.

\section{Discussion}
The maximum-likelihood (M-L) map-making method has long been used in CMB data analysis pipelines to reduce sizeable time-ordered data-sets (TOD) into optimal maps, which are optimal in the sense that the noise in the map is minimised without the loss of information - the sky signal in the maps is not distorted.  However, applying the method is moving toward being untenable for future long duration, massively multi-detector CMB experiments, establishing the need for faster, approximate, map-making methods.  Such methods will be crucial to simulating and removing experimental systematics with enough precision to search for the small primordial gravity wave signal in the CMB polarisation field.

Significant study (see eg: \citealt{ashdown:2009} and their references) has been conducted into data-analysis for space based experiments, such as the Planck experiment. We have built upon this by evaluating fast map-making methods for massively multi-detector ground based experiments, of which a large number are in development (see \citealt{brown:2009} for a review).  We have developed \textsc{descart}, an optimised, parallel destriping code, and applied it to simulations of time-ordered data (TOD) from such an experiment.  We destripe using short baselines with a noise prior - a mode of operation that has been shown to produce near optimal maps \citep{sutton:2009}.

For large future datasets, the fastest map-making method under consideration is TOD filtering.  Our comparison of the filtering and destriping approaches shows that, for the highly cross-linked scanning strategies we simulate, TOD filtering underperforms in power spectrum errors when compared to destriping.  This result is due to two effects: the suppression of signal sensitivity by the noise filter, which decreases the signal to noise ratio in the lower $\ell$ bins, despite removing the effects of correlated noise; and the introduction of $E\rightarrow B$ mixing by the TOD filter, which can be characterised and removed by TOD simulations, but which significantly contributes to the variance of the large angular scales.  Of these effects,  the latter (only present in the B-mode spectrum) is much more dominant, typically doubling the bandpower variance at the expected inflationary B-mode peak at $\ell \approx 100$, although the variance increase is dependent on how aggressive the TOD filter is.  

In our simulations, we find that bias from filtering can strongly affect the detection of the primordial B-mode peak (see Table \ref{bmode peak table}).  This is true for the one realisation of possible scan strategies, patch shapes and receiver array arrangements presented here; the full variation of the effect with the parameters of the scan and experiment is beyond the scope of this paper, and we expect that it can be ameliorated in many situations.  It is also unclear how the leakage effect will scale for longer scanning-strategies and larger focal planes.

\textsc{descart}'s computing time depends on the length of the dataset and the condition number of the destriping matrix.  Each iteration of \textsc{descart} is dominated by operations that scale linearly with the size of the dataset - the $\mathbf{P}$ and $\mathbf{F}$ operators and their transposes.  The number of iterations is sensitive to the condition of the matrix, which is generally improved by the addition of more heavily cross-linked data, such that the number of iterations required to solve for offset amplitudes reduces slightly.  For the simulations we considered, the computing time was dominated by the initial data-loading operation.  This suggests that destriping can be potentially as fast a method as filtering.

The effect of an extended focal plane is to add sufficient cross-linking to a single scan that destriping offsets are well constrained by data from that scan alone.  The gain from including multiple scans to estimate offset amplitudes is a reduction of order $0.3\%$ in map residuals, a gain that is almost entirely returned by combining a small number of scans.

The search for B-modes will likely be dependent on how well experimental and physical systematics can be constrained and mitigated.  Many of these systematics require aggressive filtering of the time-stream data prior to map-making.  However, destriping has the potential to model some of these systematic effects in addition to correlated time-stream noise.  For ground-based, constant-elevation scans, the most important systematic is scan-synchronous signal, caused for example by ground pick-up in the side-lobes of the experimental beam and time-varying atmospheric noise.  The presence of scan-synchronous signal tends to add considerably to the number of PCG iterations required to solve for offsets, as offsets have a limited capacity for modelling it. An alternative to filtering these modes is to solve for additional destriping offsets mapped by azimuth rather than by time. Such an approach is equivalent to the estimation and removal of ground signal recently applied to the QUaD data by \cite{brown:2009b}. For total power experiments, the offsets can be mapped onto all the detectors within the horn, much as atmospheric common-mode offsets are mapped onto timestreams from multiple adjacent horns in  the method of \cite{hincks:2009}.  

As part of the generalization of destriping to multiple detectors, we have included the possible effect of inter-detector correlated $1/f$ noise in the offset prior.  The use of this information in the prior results in a marginal improvement in destriping performance, but for the experimental set-up we simulate, the effect is small.  We note, however, that in other experimental situations, the inclusion of the correlation information could be important.

We have investigated the use of correlation information to constrain cross-correlated $1/f$, but have ignored the possibility of cross-correlated white noise.   Correlated white noise tends to affect detectors within the same horn, in which case it is optimally accounted for by including the correlations in the white noise covariance matrix, which we have considered to be uncorrelated here.  The detectors in the same horn observe the same sky pixels, so this extra information does not lead to replacing the naive map evaluated in each destriping iteration with a more difficult solution, as the naive mapping operation is still diagonal in pixel-space.   We plan to address experimental systematics, including correlated white noise and scan-synchronous signal, in a future paper.

\emph{Acknowledgments}: We are grateful to Bruce Winstein and the QUIET collaboration for useful discussions.  The results in this paper were obtained by making use of the HEALPix and FFTW subroutine libraries and using the computing facilities of the Research Computing Services group at the University of Oslo, and in particular the Titan cluster.  DS acknowledges the support of a Science and Technology Facilities Council PhD Studentship and JZ acknowledges support of an STFC Rolling Grant.

\end{document}